\documentclass[pra,amsmath,amssymb,amsfonts,nofootinbib]{revtex4}
\usepackage{bm,graphicx,mathrsfs}

\newcommand{\eps}{\varepsilon}

\newcommand{\tr}{\mathrm{tr}}
\newcommand{\T}{T}
\newcommand{\F}{\mathcal{F}}
\newcommand{\E}{\mathcal{E}}

\newcommand{\mc}{\mathcal}
\newcommand{\dd}{{\mathrm d}}

\newcommand{\C}{\mathscr{C}}
\newcommand{\sgn}{\mathrm{sgn}}
\newcommand{\novec}[1]{{#1}}
\newcommand{\ld}{n}
\newcommand{\qed}{}
\newcommand{\sidecaption}{}

\newtheorem{theorem}{Theorem}
\newtheorem{lemma}[theorem]{Lemma}

\newenvironment{proof}{\vspace{1.5ex}\par\noindent\textbf{Proof}}%
    {\hspace*{\fill}$\Box$\vspace{1.5ex}\par}

\newenvironment{theopargself}{}{}

\newcommand{\be}{\begin{equation}}
\newcommand{\ee}{\end{equation}}
\newcommand{\bea}{\begin{eqnarray}}
\newcommand{\eea}{\end{eqnarray}}
\newcommand{\bc}{\hspace*{\fill}}
\newcommand{\ec}{\hspace*{\fill}}

\begin{document}

\title{\sc\Large Quantum states on Harmonic lattices}
\author{Norbert Schuch} 
\affiliation{Max-Planck-Institut f\"ur Quantenoptik,
  Hans-Kopfermann-Str.\ 1, D-85748 Garching, Germany.}
\author{J.\ Ignacio Cirac}
\affiliation{Max-Planck-Institut f\"ur Quantenoptik,
  Hans-Kopfermann-Str.\ 1, D-85748 Garching, Germany.}
\author{Michael M.\ Wolf}
\affiliation{Max-Planck-Institut f\"ur Quantenoptik,
  Hans-Kopfermann-Str.\ 1, D-85748 Garching, Germany.}
\date{\today}

\begin{abstract}
We investigate bosonic Gaussian quantum states on an infinite
cubic lattice in arbitrary spatial dimensions. We derive general
 properties of such states as ground states of quadratic Hamiltonians for
both critical and non-critical cases. Tight analytic relations
between the decay of the interaction and the correlation functions
are proven and the dependence of the correlation length on band
gap and effective mass is derived. We show that properties of
critical ground states depend on the gap of the point-symmetrized
rather than on that of the original Hamiltonian. For critical
systems with polynomially decaying interactions logarithmic
deviations from polynomially decaying correlation functions are
found. Moreover, we provide a generalization of the matrix product
state representation for Gaussian states and show that properties
hold analogously to the case of finite dimensional spin systems.
\end{abstract}

\vspace*{2cm}

\maketitle

\section{Introduction}

The importance of bosonic Gaussian states arises from two facts.
First, they provide a very good description for 
accessible states of a large variety of physical systems. In fact,
every ground and thermal state of a quadratic bosonic Hamiltonian
is Gaussian and remains so under quadratic time evolutions. In
this way quadratic approximations naturally lead to Gaussian
states. Hence, they are ubiquitous in quantum optics as well as in
the description of vibrational modes in solid states, ion traps or
nanomechanical oscillators.

The second point for the relevance of Gaussian states is that they
admit a powerful phase space description which enables us to solve
quantum many-body problems which are otherwise (e.g., for spin
systems) hardly tractable. In particular, the phase space
dimension, and with it the complexity of many tasks, scales
linearly rather than exponentially in the number of involved
subsystems. For this reason quadratic Hamiltonians and the
corresponding Gaussian states also play a paradigmatic role as
they may serve as an exactly solvable toy model from which
insight into other quantum systems may be gained.

Exploiting the symplectic tools of the phase space description,
exact solutions have been found for various problems in quantum
information theory as well as in quantum statistical mechanics. In
fact, many recent works form a bridge between these two fields as
they address entanglement questions for asymptotically large
lattices of quadratically coupled harmonic oscillators: the
entropic area law \cite{PEDC05,CEPD05,Wol05} has been investigated
as well as entanglement statics \cite{AEPW02,BR04,AK05}, dynamics
\cite{PS05,PHE04,EPBH04} and frustration \cite{WVC03,WVC04}.

In the present paper we analytically derive general properties of
ground states of translationally invariant quadratic Hamiltonians
on a cubic lattice. Moreover, we provide a representation of such
states analogous to the matrix product states of finite
dimensional spin systems. We start by giving an outlook and a
non-technical summary of the main results. The results on the
asymptotic scaling of ground state correlations are summarized in
Table~\ref{table:all_scalings}.

We note that related investigations of correlation functions were
recently carried out in \cite{NS05,HK05} for finite dimensional
spin systems and in \cite{PEDC05,CE05} for generic harmonic lattices with
non-critical finite range interactions.

\clearpage

\begin{table}
\label{table:all_scalings}
\begin{tabular}{p{0.2\textwidth}||p{0.3\textwidth}|p{0.35\textwidth}}
    \bc interaction \rule[-3mm]{0mm}{8mm}\ec    &   \bc non-critical\ec     &   \bc critical\ec \\
    \hline\hline
    \centering local        &
        \begin{minipage}{0.3\textwidth}
        \centering
        \vspace{2ex}
        $\displaystyle O^*\big(e^{-n/\xi}\big)$\\[0.5ex]
            \footnotesize $d=1$:
        $\xi\sim\tfrac{1}{\sqrt{\Delta m^*}}=\tfrac v\Delta$
        \vspace{2ex}
        \end{minipage}
    & \smash{%
    \begin{minipage}{0.35\textwidth}
    \centering
    \vspace{18ex}
    \begin{tabular}{l@{\;:\quad}l}
        $d=1$ & $\displaystyle O^*\Big(\frac{1}{n^2}\Big)$\\[3ex]
        $d>1$ & $\displaystyle O\Big(\frac{\log n}{n^{d+1}}\Big)$
    \end{tabular}
    \end{minipage}
    }
    \\
    \cline{1-2}
    \begin{minipage}{0.2\textwidth}
	\centering
        \vspace{2ex}
        $O\big(n^{-\infty}\big)$
        \vspace{2ex}
    \end{minipage}
    & \centering $O\big(n^{-\infty}\big)$ & \\
    \cline{1-2}
    \begin{minipage}{0.2\textwidth}
	\centering
        \vspace{2ex}
        $\displaystyle O\Big(\frac{1}{n^\alpha}\Big)$\\[1ex]
        \footnotesize $\alpha>2d+1$
        \vspace{2ex}
    \end{minipage}
    & 
	\begin{minipage}{0.3\textwidth}
        \centering
        \vspace{2ex}
	$\displaystyle O\Big(\frac{1}{n^{\nu-d}}\Big)$\\[1.3ex]
            \footnotesize $\alpha>\nu\in\mathbb N$
        \vspace{1.7ex}
        \end{minipage}
    & \\
    \hline
    \begin{minipage}{0.2\textwidth}
	\centering
        \vspace{2ex}
        $\displaystyle\frac{c}{n^\alpha}$ \\[2ex]
        \footnotesize $d=1$
        \vspace{2ex}
    \end{minipage}
    & \begin{minipage}{0.3\textwidth}
        \centering
	$\displaystyle\alpha\ge2:\
         \displaystyle\Theta\Big(\frac{1}{n^\alpha}\Big)$
      \end{minipage} &
      \begin{minipage}{0.35\textwidth}
	\centering
        \vspace{2ex}
        \begin{tabular}{l@{\;:\quad}l}
	\centering
        $\alpha=3$&
        $\displaystyle \left\{
        \begin{array}{l@{\,,\ }l}
        \Theta\big(\frac{1}{n^2}\big) &
            c>0 \\[1ex]
        \Theta\big(\frac{\sqrt{\log n}}{n^2}\big) &
            c<0
        \end{array} \right.$\\[4ex]
         $\alpha>3$  & 
	    \hspace{1.14em}$\Theta\big(\frac{1}{n^2}\big)$ 
         \end{tabular}
        \vspace{2ex}
      \end{minipage}
\end{tabular}
\caption{Summary of the bounds derived in the paper on the
asymptotic scaling of ground state correlations, depending on the
scaling of the interaction (left column). Here $n$ is the distance
between two points (harmonic oscillators) on a cubic lattice of
dimension $d$. $O$ denotes upper bounds, $O^*$ tight upper bounds,
and $\Theta$ the exact asyptotics. The table shows the results for
generic interactions---special cases are discussed in the text.}
\end{table}

\noindent
{\bf Quadratic Hamiltonians}: In
Sec.~\ref{sec:generalities}, we
start by introducing some basic results on quadratic Hamiltonians together with the used notation. \vspace{5pt}\\
{\bf Translationally invariant systems}: In
Sec.~\ref{sec:tinv-systems}, we show first that every pure
    translational invariant Gaussian state is point symmetric.
    This implies that the spectral gap of the symmetrized rather
    than the original Hamiltonian determines the characteristic
    properties of the ground state.  We provide a general formula for
    the latter and express its covariance matrix in terms of a product of
    the inverse of the Fourier transformed
    spectral function and the Hamiltonian matrix.\vspace{5pt}\\
{\bf Non-critical systems}: Sec.~\ref{sec:noncrit} shows that if
the Hamiltonian is gapped, then
    the correlations decay according to the interaction: a (super)
    polynomial decay of the interaction leads to the same (super)
    polynomial decay for the correlations, and (following Ref.~\cite{PEDC05})
    finite range interactions lead to exponentially decaying correlations.\vspace{5pt}\\
{\bf Correlation length and gap}: Sec.~\ref{sec:corrlength-gap}
gives an explicit formula for the
    correlation length for gapped 1D-Hamiltonians with finite range
    interactions. The
    correlation length $\xi$ is expressed in terms of the dominating zero of the
    complex spectral function, which close to a critical point is in turn
    determined by the spectral gap $\Delta$ and the effective mass $m^*$ at the band gap via $\xi\sim (m^*\Delta)^{-1/2}$. When the change
    in the Hamiltonian is given by a global scaling of the interactions this proves the folk theorem $\xi\sim 1/\Delta$.\vspace{5pt}\\
{\bf Critical systems}: Sec.~\ref{sec:critical} shows that for
generic $d$-dimensional
    critical systems the correlations decay as $1/n^{d+1}$, where
    $n$ is the distance between two points on the lattice. Whereas
    for sufficiently fast decreasing interactions in
    $d=1$ the asymptotic bound is exactly polynomial, it contains an
    additional logarithmic correction for $d\geq 2$. Similarly for $d=1$ a
    logarithmic deviation is found if the interaction decays exactly like
    $-1/n^3$.
\vspace{5pt}\\
{\bf Gaussian matrix product states:} Sec.~\ref{sec:GaussianMPS}
    provides a representation of Gaussian states in terms of
    \emph{Gaussian matrix product states} (GMPS). 
    It is shown that any translational invariant pure state can be
    approximated by translational invariant GMPS to arbitrary precision, that
    the correlations of any GMPS decay exponentially, and that every GMPS
    is the ground state of a local Hamiltonian.
    \vspace{5pt}\\
 {\bf Appendix: Simulating Hamiltonians:} 
 We provide a Lemma showing that in $d=1$ every
    translational invariant Hamiltonian can be simulated by any
    translational invariant nearest neighbor Hamiltonian supplemented by
    the set of translational invariant local Hamiltonians.

\section{Quadratic Hamiltonians and their ground states
    \label{sec:generalities}}

Consider a system of $N$ bosonic modes which are characterized by
$N$ pairs of canonical operators $(Q_1,P_1,\ldots, Q_N ,P_N)=:R$.
The canonical commutation relations (CCR) are governed by the
symplectic matrix $\sigma$ via
$$ \big[R_k,R_l\big]=i\sigma_{kl}\;,\quad
\sigma=\bigoplus_{n=1}^N\left(\begin{array}{cc}0&1\\-1&0\end{array}\right)\;,
$$ and the system may be equivalently described in terms of bosonic creation and annihilation operators $a_l=(Q_l+iP_l)/\sqrt{2}$.
Quadratic Hamiltonians are of the form
$$
{\cal H}=\frac12\sum_{kl} H_{kl} R_k R_l\;,
$$
where the Hamiltonian matrix $H$ is real and positive semidefinite
due to the Hermiticity and lower semi-boundedness of the
Hamiltonian ${\cal H}$. Without loss of generality we neglect
linear and constant terms since they can easily be incorporated by
a displacement of the canonical operators and a change of the
energy offset. Before we discuss the general case we mention some
important special instances of quadratic Hamiltonians: a well
studied 1D example of this class is the case of nearest neighbor
interactions in the position operators of harmonic oscillators on
a chain with periodic boundary conditions
\begin{equation}
{\cal H}_\kappa
=\frac12\sum_{i=1}^N Q_i^2+P_i^2- \kappa\; Q_i Q_{i+1}\;,\quad
\kappa\in[-1,1]\;.
\label{eq:basics:kleingordon}
\end{equation}
This kind of spring-like interaction was studied in the context of
information transfer \cite{PS05}, entanglement statics
\cite{AEPW02,BR04,AK05} and entanglement dynamics \cite{EPBH04}.
Moreover, it can be considered as the discretization of a massive
bosonic continuum theory given by the Klein-Gordon Hamiltonian
$$
{\cal H}_{\mathrm{KG}}= \frac12
\int_{-L/2}^{L/2}
\Big[\dot{\phi}(x)^2+\big(\triangledown\phi(x)\big)^2+m^2\phi(x)^2\Big]
dx\;,
$$
where the coupling $\kappa$ is related to the mass $m$ by
$\kappa^{-1}=1+\frac12\big(\frac{mL}N\big)^2 $ \cite{BR04}. Other
finite range quadratic Hamiltonians appear as limiting cases of
finite range spin Hamiltonians via the Holstein--Primakoff
approximation \cite{Aue94}. In this way the $xy$-spin model with
transverse magnetic field can for instance be mapped onto a
quadratic bosonic Hamiltonian in the limit of strong polarization
where $a\simeq (\sigma_x+ i \sigma_y)/2$. Longer range
interactions appear naturally for instance in 1D systems of
trapped ions. These can either be implemented as Coulomb crystals
in Paul traps or in arrays of ion microtraps. When expanding
around the equilibrium positions, the interaction between two ions
at position $i$ and $j\neq i$ is---in harmonic approximation---of
the form $\frac{c\; Q_i Q_j}{|i-j|^3}$, where $c>0$ ($c<0$) if
$Q_i$, $Q_j$ are position operators in radial (axial) direction
\cite{Jam98}.

Let us now return to the general case and briefly recall the
normal mode decomposition \cite{Wil36}: every Hamiltonian matrix
can be brought to a diagonal normal form by a congruence
transformation with a symplectic matrix $S\in
\mathrm{Sp}(2N,\mathbb{R})=\{S|S\sigma
S^T=\sigma\}$:\footnote{Note that we disregard systems where the
Hamiltonian contains irrelevant normal modes.}
\begin{equation}
SHS^T=\bigoplus_{i=1}^I
    \left(\begin{array}{cc}\varepsilon_i&0\\0&\varepsilon_i\end{array}\right)
    \oplus
    \bigoplus_{j=1}^J
    \left(\begin{array}{cc}0&0\\0&1\end{array}\right),\quad
    \varepsilon_i>0\ ,
\label{eq:basics:diagonalize-H}
\end{equation}
where the \emph{symplectic eigenvalues} $\varepsilon_i$
 are the square roots of the duplicate
nonzero eigenvalues of $\sigma H\sigma^T H$. The diagonalizing
symplectic transformation $S$ has a unitary representation $U_S$
on Hilbert space which transforms the Hamiltonian according to
\begin{equation}
U_S{\cal H}U_S^\dagger = \tfrac12 \sum_{i=1}^I
\big(Q_i^2+P_i^2\big)\;\varepsilon_i + \tfrac12 \sum_{j=1}^J
P_j^2= \sum_{i=1}^I\big(a_i^\dagger
a_i+\tfrac12\big)\varepsilon_i+ \tfrac12 \sum_{j=1}^J P_j^2\;.
\label{eq:basics:hamiltonian-osc-free}
\end{equation}
Hence, by
Eq.~(\ref{eq:basics:hamiltonian-osc-free}) the ground state energy
$E_0$ and the energy gap $\Delta$ can easily be expressed in terms
of the symplectic eigenvalues of the Hamiltonian matrix:
\begin{equation}
E_0=
\tfrac12\sum_{i=1}^I\eps_i\;,\qquad
\Delta=\left\{\begin{array}{ll}
  \min_i \varepsilon_i\;,& J=0 \\
  0\;,& J> 0 \\
\end{array}\right.\;.
\label{eq:basics:gs-energy-from-sympEig}
\end{equation}
The case of a vanishing energy gap $\Delta=0$ is called
\emph{critical} and the respective ground states are often
qualitatively different from those of non-critical Hamiltonians.
For the Hamiltonian ${\cal H}_\kappa$,
Eq.~(\ref{eq:basics:kleingordon}), this happens in the strong
coupling limit $|\kappa|=1-\Delta^2\rightarrow 1$, and in the case
of 1D Coulomb crystals a vanishing energy gap in the radial modes
can be considered as the origin of a \emph{structural phase
transition} where the linear alignment of the ions becomes
unstable and changes to a zig-zag configuration
\cite{BKW92,Dub93,ESG00}. Needless to say that these phase
transitions appear as well in higher dimensions and for various
different configurations \cite{MBDHIB98}.

Ground and thermal states of quadratic Hamiltonians are
\emph{Gaussian states}, i.e, states having a Gaussian Wigner
distribution in phase space. In the mathematical physics
literature they are known as bosonic \emph{quasi-free
states}~\cite{MV68,Hol71}.
These states are completely characterized by their first moments
$d_k=\tr{\big[\rho R_k\big]}$ (which are w.l.o.g.\ set to zero in
our case) and their \emph{covariance matrix} (CM)
\begin{equation}
\gamma_{kl}=\tr\Big[\rho\big\{R_k-d_k,R_l-d_l\big\}_+\Big]\;,
\label{eq:basics:def-CM}
\end{equation}
where $\{\cdot,\cdot\}_+$ is the anticommutator. The CM satisfies
$\gamma\ge i\sigma$, which expresses Heisenberg's uncertainty
relation and is equivalent to the positivity of the corresponding
density operator $\rho\geq 0$. In order to find the ground state
of a quadratic Hamiltonian, observe that
\begin{equation}
\textstyle\frac12\displaystyle\sum_i\eps_i
\stackrel{(\ref{eq:basics:gs-energy-from-sympEig})}{=}
E_0=\inf_\rho\tr[\rho{\cal H}]
\stackrel{(\ref{eq:basics:def-CM})}{=} \textstyle
\frac14\displaystyle\inf_\gamma\tr[\gamma H]
\label{eq:basics:E0chain}\;.
\end{equation}
By virtue of
Eqs.~(\ref{eq:basics:diagonalize-H},\ref{eq:basics:hamiltonian-osc-free})
the infimum is attained for the ground state covariance matrix
\begin{equation}
\label{eq:basics:groundstate-diagonalbasis}
\gamma=\lim_{s\rightarrow\infty} S^T \left[\bigoplus_{i=1}^I
    \left(\begin{array}{cc}1&0\\0&1\end{array}\right)
    \oplus
    \bigoplus_{j=1}^J
    \left(\begin{array}{cc}s &0\\0&s^{-1}\end{array}\right)\right]S\; ,
\end{equation}
which reduces to $\gamma= S^TS$ in the non-critical case. Note
that the ground state is unique as long as $H$ does not contain
irrelevant normal modes [which we have neglected from the very
beginning in Eq.~(\ref{eq:basics:diagonalize-H})].

In many cases it is convenient to change the order of the
canonical operators such that $R=(Q_1,\ldots,Q_N,P_1,\ldots,P_N)$.
Then the covariance matrix as well as the Hamiltonian matrix can
be written in block form
$$
H=\left(\begin{array}{cc}H_Q&H_{QP}\\H_{QP}^\T&H_P\end{array}\right)\
.
$$
In this representation a quadratic Hamiltonian is particle
number preserving iff $H_Q=H_P$ and $H_{QP}=-H_{QP}^T$, that is,
the Hamiltonian  contains only terms of the kind $a_i^\dagger
a_j+a_j^\dagger a_i$. In quantum optics terms of the form
$a_i^\dagger a_j^\dagger$, which are not number preserving, are
neglected within the framework of the \emph{rotating wave
approximation}. The resulting Hamiltonians have particular simple
ground states:

\begin{theopargself}
\begin{theorem}[a]
The ground state of any particle number preserving Hamiltonian is
the vacuum with $\gamma=\openone$, and the corresponding ground
state energy is given by $E_0=\frac14\tr H$.
\end{theorem}
\end{theopargself}

\begin{proof}
Number preserving Hamiltonians are most easily expressed in terms
of creation and annihilation operators. For this reason we change
to the respective complex representation via the transformation
$$
H\ \mapsto\ \Omega H\Omega^T=
    \left(\begin{array}{cc}0& X\\ \bar{X}&0
    \end{array}\right)\;,\quad
     \Omega=\frac{1}{\sqrt{2}}
    \left(\begin{array}{cc}\openone&-i\openone\\
        \openone&i\openone\end{array}\right)\;.
$$
In this basis $H$ is transformed to normal form via a block
diagonal unitary transformation  $U\oplus\bar U$ which in turn
corresponds to an element of the orthogonal subgroup of the
symplectic group $\mathrm{Sp}(2N,\mathbb{R})\cap
\mathrm{SO}(2N)\simeq \mathrm{U}(N)$ \cite{ADMS95b}. Hence, the
diagonalizing $S$ in
Eqs.~(\ref{eq:basics:diagonalize-H},\ref{eq:basics:groundstate-diagonalbasis})
is orthogonal and since $J=0$ due to particle number conservation,
we have $\gamma=S^TS=\openone$. $E_0$ follows then immediately
from Eq.~(\ref{eq:basics:E0chain}).
\hspace*{\fill}\qed
\end{proof}

Another important class of quadratic Hamiltonians for which the
ground state CM takes on a particular simple form corresponds to
the case $H_{QP}=0$ where there is no coupling between the
momentum and position operators:
\setcounter{theorem}0
\begin{theopargself}
\begin{theorem}[b] For a quadratic
Hamiltonian with Hamiltonian matrix $H=H_Q\oplus H_P$ the ground
state energy and the ground state CM are given by 
\begin{equation}
E_0=\tfrac12\tr\big[\sqrt{H_Q}\sqrt{H_P}\big],\
\gamma=X\oplus X^{-1},\ 
X=H_Q^{-1/2}\sqrt{H_Q^{1/2}H_P H_Q^{1/2}}H_Q^{-1/2}.
\label{eq:basics:noQP}
\end{equation}
\end{theorem}
\end{theopargself}

\begin{proof}
Since $\sigma H\sigma^TH= H_PH_Q\oplus H_QH_P$, the symplectic
eigenvalues of $H$ are given by the eigenvalues of
$\sqrt{H_Q}\sqrt{H_P}$ and thus
$E_0=\tfrac12\tr\big[\sqrt{H_Q}\sqrt{H_P}\big]$. Moreover, by the
uniqueness of the ground state and the fact that
$E_0=\frac14\tr[\gamma H]$ with $\gamma$ from
Eq.~(\ref{eq:basics:noQP}) we know that $\gamma$ is the ground
state CM (as it is an admissible pure state CM by construction).
\hspace*{\fill}\qed
\end{proof}
Finally we give a general formula for the ground state CM in cases
where the blocks in the Hamiltonian matrix can be diagonalized
simultaneously. This is of particular importance as it applies to
all translational invariant Hamiltonians discussed in the
following sections.
\setcounter{theorem}0
\begin{theopargself}
\begin{theorem}[c]
Consider a quadratic Hamiltonian for which the blocks
$H_Q$, $H_P$, $H_{QP}$ of the Hamiltonian matrix can be diagonalized
simultaneously and in addition $H_{QP}=H_{QP}^T$. Then with
\begin{eqnarray}
\hat\E&=&\sqrt{H_Q H_P-H_{QP}^2}\quad\quad\text{we have}
\label{eq:basics:tinv-Ehat-definition}\\
E_0&=&\tfrac12\tr[\hat\E]\;,\quad
\Delta=\lambda_{\min}\big(\hat\E\big)\;,\quad
\gamma=(\hat\E\oplus\hat\E)^{-1}\sigma H\sigma^T\;.
\label{eq:basics:tinv-E0-and-gamma_from_Ehat}
\end{eqnarray}
\end{theorem}
\end{theopargself}

\begin{proof}
Since $\sigma H\sigma^TH= \hat\E^2\oplus\hat\E^2$ we have indeed
$E_0=\frac12\tr[\hat\E]$ and
$\Delta=\lambda_{\min}\big(\hat\E\big)$. Positivity $\gamma\geq 0$
is implied by $H\geq 0$ such that we can safely talk about the
symplectic eigenvalues of $\gamma$. The latter are, however, all
equal to one due to $(\gamma\sigma)^2=-\openone$ so that $\gamma$
 is an admissible pure state CM. Moreover it belongs to the
ground state since $\frac14\tr[H\gamma]=E_0$.
\hspace*{\fill}\qed
\end{proof}

\section{Translationally invariant systems
    \label{sec:tinv-systems}}

Let us now turn towards translationally invariant systems. We
consider cubic lattices in $d$ dimensions with periodic boundary
conditions. For simplicity we assume that the size of the lattice
is $N^d$.
The system is again characterized by a Hamiltonian matrix
$H_{kl}$, where the indices $k,l$, which correspond to two points
(harmonic oscillators) on the lattice, are now $d$-component
vectors in $\mathbb{Z}_N^d$. Translational invariance is then
reflected by the fact that any matrix element $A_{kl}$,
$A\in\{H_Q,H_P,H_{QP}\}$ depends only on the relative position
$k-l$ of the two points on the lattice, and we will therefore
often write $A_{k-l}=A_{kl}$. Note that due to the periodic
boundary conditions $k-l$ is understood modulo $N$ in each
component. Matrices of this type are called \emph{circulant}, and
they are all simultaneously diagonalized via the Fourier transform
\begin{eqnarray*}
\F_{\alpha\beta}&=&\frac1{\sqrt{N}}\;
    e^{\frac{2\pi i}N\alpha\beta}\; ,\quad
    \alpha,\beta\in\mathbb{Z}_N\;,\qquad\text{such that}\\
\hat A&:=&\F^{\otimes d}A\F^{\dagger\otimes d}=
\mathrm{diag}\left(\sum_{n\in \mathbb{Z}_N^d}
    A_{\novec n}\;e^{-\frac{2\pi i}{N} m\, n}\right)_{\novec m}\;
    ,
\end{eqnarray*}
where $m\, n$ is the usual scalar product in $\mathbb{Z}_N^d$. It
follows immediately that all circulant matrices mutually commute.

In the following, we will show that we can without loss of
generality restrict ourselves to point-symmetric Hamiltonians,
i.e., those for which $H_{QP}=H_{QP}^\T$ (which means that $\cal
H$ contains only pairs $Q_k P_l+Q_l P_k$). For dimension $d=1$
this is often called reflection symmetry.

\begin{theorem}
\label{theorem:pure-reflection-symmetry} Any translationally
invariant pure state CM $\Gamma$ is point symmetric.
\end{theorem}

\begin{proof}
For the proof, we use that any pure state covariance
matrix can be written as
$$
\Gamma=\left(\begin{array}{cc}
    \Gamma_Q&\Gamma_{QP}\\\Gamma_{QP}^\T&\Gamma_P\end{array}\right)
=\left(\begin{array}{cc}X&XY\\YX\ &\ X^{-1}+YXY\end{array}\right)\ ,
$$
where $X\ge0$ and $Y$ is real and symmetric~\cite{WGKWC03}. From
translational invariance, it follows that all blocks and thus $X$
and $Y$ have to be circulant and therefore commute. Hence,
$\Gamma_{QP}=XY=YX=\Gamma_{QP}^T$, i.e., $\Gamma$ is point
symmetric.
\hspace*{\fill}\qed
\end{proof}
Let ${\cal P}:\mathbb{Z}_N^d\rightarrow \mathbb{Z}_N^d$ be the
reflection on the lattice and define the symmetrization operation
${\cal S}(A)=\frac12(A+{\cal P}A{\cal P})$ such that by the above
theorem ${\cal S}(\gamma)=\gamma$ for every translational
invariant pure state CM. Then due to the cyclicity of the trace we
have for any translational invariant Hamiltonian
$$
\inf_\gamma\tr\big[H\gamma\big]=\inf_\gamma\tr\big[{\cal
S}(H)\gamma\big]\;.
$$
Hence, the point-symmetrized Hamiltonian
${\cal S}(H)$, which differs from $H$ by the off-diagonal block
${\cal S}(H_{QP})=\frac12(H_{QP}+H_{QP}^T)$ has both the same
ground state energy and the same ground state as $H$. Together
with Theorem~1c this leads us to the following:

\begin{theorem}\label{thm:symH}
Consider any translationally invariant quadratic Hamiltonian. With
$\hat\E=\big[H_Q H_P- \tfrac14(H_{QP}+H_{QP}^T)^2\big]^{1/2}$ the
ground state CM and the corresponding ground state energy are
given by
\begin{equation}
E_0=\tfrac12\tr[\hat\E]\;,\quad
\gamma=\big(\hat\E\oplus\hat\E\big)^{-1}\sigma{\cal
S}(H)\sigma^T\;.\label{eq:basics:symresult}
\end{equation}
\end{theorem}

It is important to note that the energy gaps of $H$ and ${\cal
S}(H)$ will in general be different. In particular $H$ might be
gapless while ${\cal S}(H)$ is gapped. However, as we will see in
the following sections, the properties of $\gamma$ depend on the
gap $\Delta=\lambda_{\min}(\hat\E)$ of the symmetrized Hamiltonian
rather than on that of the original $H$. For this reason we will
in the following for simplicity assume $H_{QP}=H_{QP}^T$. By
Thm.\;3 all results can then also be applied to the general case
without point symmetry if one only keeps in mind that $\Delta$ is
the gap corresponding to ${\cal S}(H)$.

Note that the eigenvalues of $\hat\E$ are the symplectic
eigenvalues of ${\cal S}(H)$, i.e., $\E=\F^{\otimes
d}\hat\E\F^{\dagger\otimes d}$ is the excitation spectrum of the
Hamiltonian. This is the reason for the notation where $\E$
resides in Fourier space and $\hat\E$ in real space, which is
differs from the normal usage of the hat  throughout the
paper.\vspace{5pt}\\
{\bf Correlation functions.} According to 
Eqs.~(\ref{eq:basics:tinv-Ehat-definition},%
\ref{eq:basics:tinv-E0-and-gamma_from_Ehat},\ref{eq:basics:symresult})
we have to compute the entries of functions of matrices in order
to learn about the entries of the covariance matrix. This is most
conveniently done by a double Fourier transformation, where one
uses that $\widehat{f(M)}=f(\hat M)$, and we find\vspace{-5pt}
\begin{equation}
[f(M)]_{\novec n\novec m}=\frac1{N^d}\sum_{r,s}
    e^{-\frac{2\pi i}{N} \, n r}
   [f(\hat M)]_{rs}e^{\frac{2\pi i}{N}s m}\ .
\label{eq:corr:FT-1}
\end{equation}
As we consider translationally invariant systems, $M$ is circulant
and thus $\hat M$ is diagonal. We define the function\vspace{-5pt}
\begin{equation}
\label{eq:corr:def-finite-hat-M} \hat
M(\novec\phi)=\sum_{\ld\in\mathbb Z^d_N}
    M_{\novec \ld}\,e^{-i\novec \ld\novec \phi}
\end{equation}
such that $\hat M(2\pi\novec r/N)=\hat M_{\novec r,\novec r}$. As
$f(M)$ is solely determined by its first row, we can write \be
\label{guglhupf} [f(M)]_n=
    \frac{1}{N^d}\sum_{r\in\mathbb Z_N^d}
    e^{2\pi i\, \novec \ld \novec r/N}f(\hat M(2\pi\novec r/N))\; .
\ee In the following we will use the index $n\in\mathbb{Z}^d$ for
the relative position of two points on the lattice. Their distance
will be measured either by the $l_1,l_2$ or $l_\infty$ norm. Since
we are considering finite dimensional lattices these are all
equivalent for our purpose and we will simply write $\|n\|$. In
the thermodynamic limit $N\rightarrow\infty$, the sum in
Eq.~(\ref{guglhupf}) converges to the integral
\begin{equation}
\label{eq:corr:def-hat-M} [f(M)]_n=\frac{1}{(2\pi)^d}\int_{\mc
T^d}\mathrm{d}\phi\,
    f(\hat M(\novec\phi))\;e^{i\novec \ld\novec\phi}\quad\text{with}\quad \hat M(\novec\phi)=\sum_{\novec
\ld\in\mathbb Z^d}
    M_{\novec \ld}\,e^{-i\novec \ld\novec \phi}\ ,
\end{equation}
where $\mc T^d$ is the $d$-dimensional torus, i.e.,  $[0,2\pi]^d$
with periodic boundary conditions. The convergence holds as soon as
$\sum |M_n|<\infty$ [which holds e.g.\ for
$M_n=O(\|n\|^{-\alpha})$ with some $\alpha>d$] and $f$ is continuous on 
an open interval which contains the range of $\hat M$. 

From the definition (\ref{eq:corr:def-hat-M}) of $\hat M$, it
follows that $\hat M\in\C^k(\mc T^d)$ (the $n$ times continuously
differentiable functions on $\mc T^d$) whenever the entries
$M_{\ld}$ decay at least as fast as $\|\ld\|^{-\alpha}$ for some
$\alpha>k+d$, since then the sum of the derivatives converges
uniformly. Particularly, if the entries of $M$ decay faster than any
polynomial, then $\hat M\in\C^\infty(\mc T^d)$. In the following
the most important function of the type $f\circ\hat M$ will be the
\emph{spectral function} \be\E(\phi)=\sqrt{\sum_{n\in\mathbb{Z}^d}
e^{-i n \phi}\Big([H_Q H_P]_n-[H_{QP}^2]_n\Big)
}\label{eq:spectralfunction}\;.\ee {\bf Asymptotic notation.} As
the main issue of this paper is the asymptotic scaling of
correlations, we use the Landau symbols $o$, $O$, and $\Theta$, as
well as the symbol $O^*$ for tight bounds:
\begin{itemize}
\item $f(x)=o(g(x))$ means
    $\lim\limits_{x\rightarrow\infty}\tfrac{f(x)}{g(x)}=0$, i.e., $f$
    vanishes strictly faster than $g$ for $x\rightarrow\infty$;
\item $f(x)=O(g(x))$, if $\limsup\limits_{x\rightarrow\infty}
\left|\tfrac{f(x)}{g(x)}\right|$ is finite, i.e., $f$ vanishes at least as
fast as $g$;
\item $f(x)=\Theta(g(x))$, if $f(x)=O(g(x))$ and $g(x)=O(f(x))$ (i.e.,
exact asymptotics); \item $f(x)=O^*(g(x))$, if $f(x)=O(g(x))$ but $f(x)\ne
o(g(x))$, i.e., $g$ is a tight bound on $f$.\footnote{
    In order to see the difference to $\Theta$, take an $f(x)=g(x)$ for
    even $x$, $f(x)=0$ for odd $x$, $x\in\mathbb N$. Although $f$ does not
    bound $g$, thus $f(x)\ne O(g(x))$, the bound $g$ is certainly tight.
    A situation like this is met, e.g., in
    Theorem~\ref{theorem:noncrit-1D-corrlength}, where the correlations
    oscillate within an exponentially decaying envelope.} If $f$ is taken
    from a set (e.g., those function consistent with the assumptions of a
    theorem) we will write $f=O^*(g)$ if $g$ is a tight bound for at least
    one $f$ (i.e., the best possible universal bound under the given
    assumptions).
\end{itemize}
If talking about Hamiltonians, the scaling is meant to hold for
all blocks, e.g., if the interaction vanishes as
$O(\|n\|^{-\alpha})$ for $n\rightarrow\infty$, this holds for all
the blocks $H_Q$, $H_P$, and $H_{QP}=H_{PQ}^T$. The same holds for
covariance matrices in the non-critical case. By the shorthand
notation $f(n)=o(\|\ld\|^{-\infty})$, we mean that
$f(n)=o(\|n\|^{-\alpha})\ \forall\alpha>0$.  Note finally that the
Landau symbols are also used in (Taylor) expansions around a point
$x_0$ where the considered limit is $x\rightarrow x_0$ rather than
 $x\rightarrow\infty$.

\section{Non-critical systems
    \label{sec:noncrit}}

In this section, we analyze the ground state correlations of
non-critical systems, i.e., those which exhibit an energy gap
$\Delta>0$ between the ground an the first excited state. Simply
speaking, we will show that the decay of correlations reflects the decay
of the interaction. While local (super-polynomially decaying) interactions
imply exponentially (super-polynomially) decaying correlations, a
polynomial decay of interactions will lead to the same polynomial law for
the correlations.

According to Theorem~\ref{thm:symH}, we  will consider a translationally
invariant system with a point-symmetric Hamiltonian
($H_{QP}=H_{QP}^T$). Following
(\ref{eq:basics:tinv-E0-and-gamma_from_Ehat},\ref{eq:basics:symresult}),
we have to determine the entries of
$(\hat\E^{-1}\oplus\hat\E^{-1})\sigma H\sigma^T$, with
$\hat\E=(H_QH_P+H_{QP}^2)^{1/2}$.  In
Lemma~\ref{lemma:poly-convolution} we will first show that it is
possible to consider the two contributions independently, and as
the asymptotics of $\sigma H\sigma^T$ is known, we only have to
care about the entries of $\hat \E^{-1}$, i.e., we have to
determine the asymptotic behavior of the integral
$$
(\hat\E^{-1})_n=\frac{1}{(2\pi)^d}\int_{\mc T^d}\mathrm{d}\phi\,
    \E^{-1}(\novec\phi)e^{i\novec
    \ld\novec\phi}\quad\text{where}\quad \E=(\hat H_Q\hat H_P+\hat
    H_{QP}^2)^{1/2}\;.
$$

\begin{lemma}
\label{lemma:poly-convolution}
Given two asymptotic circulant matrices $A$, $B$ in $d$ dimensions
with polynomially decaying entries,
$A_{\novec \ld}=O(\|\novec \ld\|^{-\alpha})$,
$B_{\novec \ld}=O(\|\novec \ld\|^{-\beta})$, $\alpha,\beta>d$.
Then
$$
(AB)_{\novec \ld}=O^*(\|\novec \ld\|^{-\mu})\ ,\quad
    \mu:=\min\{\alpha,\beta\}\;.
$$
\end{lemma}

\begin{proof} With $Q_\eta(\ld):=\min\{1,\|\ld\|^{-\eta}\}$,
we know that $|A_{\novec \ld}| = O( Q_\alpha)$ and $|B_{\novec
\ld}|=O(Q_\beta)$, and
\begin{equation}
|(AB)_n|=\left|\sum_jA_{0,\novec j}B_{\novec j,\novec
\ld}\right|\le \sum_j|A_{\novec j}||B_{\novec \ld-\novec j}|
=O\Big( \sum_j Q_\alpha(\novec j)Q_\beta(\novec \ld-\novec
j)\Big)\ . \label{eq:noncrit:poly-convolution}
\end{equation}
We consider only one half space $\|\novec j\|\le\|\novec
\ld-\novec j\|$, where we bound $Q_\beta(\novec \ld-\novec j)\le
Q_\beta(\novec \ld/2)$. As $Q_\alpha(\novec j)$ is summable, the
contribution of this half-plane is  $O\big(Q_\beta(\novec
\ld/2)\big)$. The other half-plane gives the same result with
$\alpha$ and $\beta$ interchanged, which proves the bound, while
tightness follows by taking all $A_n$, $B_n$ positive.
\hspace*{\fill}\qed
\end{proof}

We now determine the asymptotics of $(\hat \E^{-1})_n$ for different
types of Hamiltonians.

\begin{lemma}
\label{lemma:rapid-H-rapid-E}
For non-critical systems with rapidly decaying interactions, i.e.,
as $o(\|\novec \ld\|^{-\infty})$, the
entries of $\hat \E^{-1}$ decay rapidly as well. That is,
$$
\Delta>0\ \Rightarrow\
    (\hat\E^{-1})_n=o(\|\novec \ld\|^{-\infty})\ .
$$
\end{lemma}

\begin{proof}
As the interactions decay as $o(\|\novec \ld\|^{-\infty})$, $\hat
H_{\bullet}\in\C^\infty(\mc T^d)$ ($\bullet=Q,P,PQ$), and thus
$\E^2=\hat H_Q\hat H_P+\hat H_{QP}^2\in\C^\infty(\mc T^d)$. Since
the system is gapped, i.e., $\E\ge\Delta>0$, it follows that also
$g:=\E^{-1}\in\C^\infty(\mc T^d)$. For the proof, we need to bound
$$
(\hat \E^{-1})_n=\frac{1}{(2\pi)^d}\int_{\mc T^d}\mathrm{d}\phi\,
    g(\novec\phi)e^{i\novec \ld\novec\phi}
$$
by $\|n\|^{-\kappa}$ for all $\kappa\in\mathbb N$.
First, let us have a look at the one-dimensional case.
By integration by parts, we get
$$
(\hat \E^{-1})_n=
  \frac{1}{2\pi}
    \left[\frac{1}{i\ld}g(\phi)e^{i\ld\phi}\right]_{\phi=-\pi}^{\pi}
  -\frac{1}{2\pi i\ld}\int_{-\pi}^{\pi}\mathrm d\phi\,
    g'(\phi)e^{i\ld\phi}\ ,
$$
where the first part vanishes due to the periodicity of $g$. As
$g\in\C^\infty(\mathcal T^1)$, the integration by parts can be iterated
arbitrarily often and all the brackets vanish, such that after $\kappa$
iterations,
$$
(\hat \E^{-1})_n=
    \frac{1}{2\pi(i\ld)^\kappa}\int_{-\pi}^{\pi}\mathrm d\phi\,
    g^{(\kappa)}(\phi)e^{i\ld\phi}\ .
$$
As $g^{(\kappa)}(\phi)$ is continuous,
the integral can be bounded by $\int|g^{(\kappa)}(\phi)|\dd\phi=:
C_\kappa<\infty$,
such that finally
$$
|(\hat \E^{-1})_n|\le
    \frac{C_\kappa}{\ld^\kappa}\quad\forall \kappa\in\mathbb N\,,
$$
which completes the proof of the one-dimensional case.

The extension to higher dimensions is straightforward. For a
given $\novec \ld=(\ld_1,\dots,\ld_d)$, integrate by parts with
respect to the $\phi_i$ for which $|\ld_i|=\|\ld\|_\infty$; we
assume $i=1$ without loss of generality. As
$g(\cdot,\phi_2,\dots,\phi_d)\in \C^\infty(\mc S^1)$, the same
arguments as in the 1D case show
$$
|(\hat\E^{-1})_n|\le \frac{1}{(2\pi)^d|\ld_1|^\kappa}
    \int_{\mc T^d}
    \left|\frac{\partial^\kappa}{\partial\phi_1^\kappa}
    g(\novec\phi)\right|\dd\novec\phi
=\frac{C_\kappa}{\|\ld\|_\infty^\kappa}\ .
$$
\hspace*{\fill}\qed
\end{proof}

For systems with \emph{local} interactions, a stronger version of
Lemma \ref{lemma:rapid-H-rapid-E} can be obtained:

\begin{lemma}
\label{lemma:finite-H-exp-E}
For a system with finite range interaction, the entries of
$\hat\E^{-1}$ decay exponentially.
\end{lemma}

This has been proven in~\cite{PEDC05} for Hamiltonians of the
type $H=V\oplus\openone$, exploiting a result on functions of
banded matrices~\cite{BG99}. Following 
Eqs.~(\ref{eq:basics:tinv-Ehat-definition},%
\ref{eq:basics:symresult})
the generalization to arbitrary translational invariant
Hamiltonians is straightforward by replacing $V$ with
$H_QH_P-H_{QP}^2$. In fact, it has been shown recently that the
result even extends to non translational invariant Hamiltonians of the
form in Theorem~1 b \cite{CE05}.

Finally, we consider systems with polynomially decaying interaction.

\begin{lemma}
\label{lemma:poly-int-poly-decay}
For a 1D lattice with $H=V\oplus\openone>0$ and an
exactly polynomially decaying interaction
$$
V_{ij}=\begin{cases}
    i=j&:\quad a\\
    i\ne j&:\quad \frac{b}{|i-j|^\nu}\
\end{cases}\ ,\ 2\leq\nu\in{\mathbb N}\;,
$$
$\hat\E^{-1}$ decays polynomially with the same exponent,
$(\hat\E^{-1})_{n}=(V^{1/2})_n=\Theta(|n|^{-\nu})$.
\end{lemma}

Hamiltonians of this type appear, e.g., for the
vibrational degrees of freedom of ions in a linear trap, where
$\nu=3$.

\begin{proof}
We need to estimate
$
(\hat\E^{-1})_n
\stackrel{(\ref{eq:basics:tinv-Ehat-definition})}{=}
(V^{-1/2})_n=
    \frac{1}{2\pi}
    \int_{0}^{2\pi}{\hat V^{-1/2}(\phi)}e^{i\ld\phi}\dd\phi\ .
$ Note that
\begin{equation}
\label{eq:noncrit:poly:LiDef}
\hat V(\phi)=a+2b\sum_{\ld=1}^\infty\frac{\cos(\ld\phi)}{\ld^\nu}=
    a+2b\,\mathrm{Re}\left[\mathrm{Li}_\nu(e^{i\phi})\right]>0\ ,
\end{equation}
where $\mathrm{Li}_\nu(z)=\sum_{n\ge1}z^n/n^\nu$ is the polylogarithm.
The polynomial decay of coefficients implies
$\hat V\in\mathscr C^{\nu-2}(\mc S^1)$, and
as the system is non-critical,
$\hat V^{-1/2}\in\mathscr C^{\nu-2}(\mc S^1)$.
As $\mathrm{Li}_\nu$ has an analytic continuation to
$\mathbb C\backslash[1;\infty)$,
$\hat V\in\C^\infty((0;2\pi))$ and thus $\hat V^{-1/2}\in
\C^\infty((0;2\pi))$.
We can therefore integrate by parts $\nu-1$ times,
and as all brackets vanish due to periodicity, we obtain
\begin{equation}
(\hat\E^{-1})_n=
    \frac{1}{2\pi(i\ld)^{\nu-1}}\int_{0}^{2\pi}
    \left[\frac{\dd^{\nu-1}}{\dd\phi^{\nu-1}}
    {\hat V^{-1/2}(\phi)}\right]
        e^{i\ld\phi}\dd\phi\ ,
\label{eq:noncrit:poly:n-1-partial}
\end{equation}
and
\begin{equation}
\label{eq:noncrit:poly-int-n-1-partial-terms}
\frac{\dd^{\nu-1}}{\dd\phi^{\nu-1}}{\hat V^{-1/2}(\phi)}=
-\frac{\hat V^{(\nu-1)}(\phi)}{2\hat V(\phi)^{3/2}}+
\frac{3(\nu-2)\hat V^{(\nu-2)}(\phi)\hat V^{(1)}(\phi)}
    {4\hat V(\phi)^{5/2}}+
g(\phi)\ .
\end{equation}
Note that the second term only appears if $\nu\ge3$, and $g$ only if $\nu\ge4$.
As $g(\phi)\in\mathscr C^1(\mc S^1)$, its  Fourier coefficients vanish as
$O(\ld^{-1})$, as can be shown by integrating by parts. The second term can
be integrated by parts as well, the bracket vanishes due to continuity,
and we remain with
$$
\frac{1}{i\ld}\int_0^{2\pi}\left[
\frac{3(\nu-2)\hat V^{(\nu-1)}(\phi)\hat V^{(1)}(\phi)}
    {4\hat V(\phi)^{5/2}}
+h(\phi)\right]e^{i\ld\phi}\dd\phi\ ,
$$
with $h\in\mathscr C(\mc S^1)$. [For $\nu=3$, a factor $2$ appears
as $(\hat V^{(1)})'=\hat V^{(\nu-1)}$.] As we will show later,
$\hat V^{(\nu-1)}$ is absolutely integrable, hence the integral
exists, and thus the Fourier coefficients of the second term in
Eq.~(\ref{eq:noncrit:poly-int-n-1-partial-terms}) vanish as
$O(n^{-1})$ as well. Finally, it remains to bound
\begin{equation}
\label{eq:noncrit:poly:lastintegral}
\int_0^{2\pi}
\frac{\hat V^{(\nu-1)}(\phi)}{2\hat V(\phi)^{3/2}}e^{i\ld\phi}\dd\phi\ .
\end{equation}
As $\mathrm{Li}_\nu'(x)=\mathrm{Li}_{\nu-1}(x)/x$, it follows from
Eq.~(\ref{eq:noncrit:poly:LiDef}) that
\begin{eqnarray*}
V^{(\nu-1)}(\phi)&=&
    2b\,\mathrm{Re}\left[i^{\nu-1}\mathrm{Li}_1(e^{i\phi})\right]
\;=\;2b\,\mathrm{Re}\left[-i^{\nu-1}\log(1-e^{i\phi})\right]\ ,
\end{eqnarray*}
where the last step is from the definition of $\mathrm{Li}_1$.

We now distinguish two cases. First, assume that $\nu$ is even. Then,
$$
V^{(\nu-1)}(\phi)\propto\mathrm{Im}\log(1-e^{i\phi})
    =\arg(1-e^{i\phi})=(\phi-\pi)/2\
$$
on $(0;2\pi)$, hence the integrand in
Eq.~(\ref{eq:noncrit:poly:lastintegral}) is bounded and has a
bounded derivative, and by integration by parts, the integral
Eq.~(\ref{eq:noncrit:poly:lastintegral}) is $O(\ld^{-1})$. In case
$\nu$ is odd we have
\begin{eqnarray*}
V^{(\nu-1)}(\phi)&\propto&\mathrm{Re}\log(1-e^{i\phi})
    \;=\;\log\left|1-e^{i\phi}\right|=\log(2\sin(\phi/2))
\end{eqnarray*}
on $(0;2\pi)$. With $h(\phi):=\hat V^{-3/2}(\phi)/2$, the
integrand in Eq.~(\ref{eq:noncrit:poly:lastintegral}) can be
written as
\begin{equation}
\hat V^{(\nu-1)}(\phi)h(\phi)\propto
    \log(2\sin(\phi/2))\;h(0)+
    \log(2\sin(\phi/2))\;[h(\phi)-h(0)]\ .
\label{eq:noncrit:poly:the-h-decomposition}
\end{equation}
The first term gives a contribution proportional to
$$
\int_{0}^{2\pi}\log(2\sin(\phi/2))\cos(\ld\phi)\dd\phi
=-\frac{1}{2\ld}
$$
as it is the back-transform of $-\frac12\sum_{\ld\ge1}\cos(\ld\phi)/\ld$.
For the second term, note that $h\in\mathscr C^1(\mc S^1)$ for $\nu\ge3$
and thus $h(\phi)-h(0)=h'(0)\phi+o(\phi)$ by Taylor's theorem.
Therefore, the $\log$ singularity vanishes, and we can once more
integrate by parts.  The derivative is
$$
\frac{1}{2}\cot(\phi/2)\;[h(\phi)-h(0)]+
    \log(2\sin(\phi/2))\;h'(\phi)\ .
$$
In the left part, the $1/\phi$ singularity of $\cot(\phi/2)$
is cancelled out by $h(\phi)-h(0)=O(\phi)$, and
the second part is integrable as $h'\in\mathscr C(\mc S^1)$, so that the
contribution of the integral (\ref{eq:noncrit:poly:lastintegral}) is
$O(\ld^{-1})$ as well.

In order to show that $n^{-\nu}$ is also a lower bound on $(\hat
V^{-1/2})_n$, one has to analyze the asymptotics more carefully.
Using the Riemann-Lebesgue lemma---which says that the Fourier
coefficients of absolutely integrable functions are $o(1)$---one
finds that all terms in (\ref{eq:noncrit:poly:n-1-partial}) vanish
as $o(1/n^\nu)$, except for the integral
(\ref{eq:noncrit:poly:lastintegral}). Now for even $\nu$,
(\ref{eq:noncrit:poly:lastintegral}) can be integrated by parts,
and while the brackets give a $\Theta(n^{-\nu})$ term, the
remaining integral is $o(n^{-\nu})$, which proves that $(\hat
V^{-1/2})_n=\Theta(n^{-\nu})$. For odd $\nu$, on the other hand,
the first part of (\ref{eq:noncrit:poly:the-h-decomposition})
gives exactly a polynomial decay, while the 
contributions from the second part vanishes
as $o(n^{-\nu})$, which proves $(\hat
V^{-1/2})_n=\Theta(n^{-\nu})$ for odd $\nu$ as well.
\hspace*{\fill}\qed
\end{proof}

\textbf{Generalizations of Lemma~\ref{lemma:poly-int-poly-decay}.} 
The preceding lemma can be extended to
non-integer exponents $\alpha\not\in\mathbb N$:
if $V_n\propto n^{-\alpha}$, $n\ne0$, then 
$(\hat\E^{-1})_n=O(n^{-\alpha})$.

For the proof, define $\alpha=\nu+\eps$, $\nu\in\mathbb N$, $0<\eps<1$. 
Then $\hat V\in\C^{\nu-1}(\mc S^1)$, $\hat V\in\C^\infty((0;2\pi))$,
and one can integrate by parts $\nu$ times, where all brackets vanish. 
What remains is to bound the Fourier integral of the $\nu$'th derivative
of $\hat V^{-1/2}$ by $n^{-\eps}$. An upper bound can be established by 
noting that 
$|\hat V^{(\nu)}(\phi)|\le|\mathrm{Li}_\eps(e^{i\phi})|=O(\phi^{\eps-1})$ 
and $|\hat V^{(\nu+1)}(\phi)|=O(\phi^{\eps-2})$. It follows that
all contributions in the Fourier integral except the singularity from
$\hat V^{(\nu)}$ lead to $o(1/n)$ contributions as can be shown by 
another integration
by parts. In order to bound the Fourier integral of the
$O(\phi^{\eps-1})$ term, split the Fourier integral at $\tfrac1n$. The
integral over $[0;\tfrac1n]$ can be directly bounded by $n^{-\eps}$,
while for $[\tfrac1n;1]$, an equivalent bound can be established after
integration by parts, using $\hat V^{(\nu+1)}=O(\phi^{\eps-2})$.
This method is discussed in more detail in the proof of
Theorem~\ref{theorem:critical-multidim}, following
Eq.~(\ref{eq:crit:g-fourier-int}).

The proof that $n^{-\eps}$ is also a lower bound to $(\hat\E^{-1})_n$ is
more involved.  From a series expansion
of $\hat V$ and its derivatives, it can be seen that it suffices to bound
the sine and cosine Fourier coefficients of $\phi^{\eps-1}$ from below.
As in the proof of Theorem~\ref{theorem:crit:iontrap-logcorrection},
this is accomplished by splitting the integral into single oscillations 
of the sine or cosine and bounding each part by the derivative of
$\phi^{\eps-1}$. 

For polynomially bounded interactions $V_n=O(n^{-\alpha})$, $\alpha>1$, not
very much can be said without further knowledge. With $\nu<\alpha$,
$\nu\in\mathbb N$  the largest integer strictly smaller than $\alpha$, we
know that $\hat V\in\C^{\nu-1}(\mc S^1)$. Thus, one can integrate by parts
$\nu-1$ times, the brackets vanish, and the remaining Fourier integral is
$o(1)$ using the Riemann-Lebesgue lemma. It follows that 
$(\hat\E^{-1})_n=o\big(n^{-(\nu-1)}\big)$. In contrast to the case of an
exactly polynomial decay, this can be extended to higher spatial
dimensions $d>1$ by replacing $\nu-1$ with $\nu-d$, which yields 
$(\hat\E^{-1})_n=o\big(n^{-(\nu-d)}\big)$.

We now use the preceding lemmas about the entries of $\hat \E^{-1}$ 
(Lemma~\ref{lemma:rapid-H-rapid-E}--\ref{lemma:poly-int-poly-decay}) to
derive corresponding results on the correlations of ground states
of non-critical systems.

\hspace{-1.3em}\begin{minipage}{\textwidth}
\begin{theorem}
\label{theorem:noncrit-decay}
For systems with $\Delta>0$, the following holds:
\begin{itemize}
\item[\textit{(i)}] If the Hamiltonian $H$ has finite range, the
ground state correlations decay exponentially.
\item[\textit{(ii)}] If $H$ decays as $o(\|\novec \ld\|^{-\infty})$, the
ground state correlations decay as $o(\|\novec \ld\|^{-\infty})$ as well.
\item[\textit{(iii)}] For a 1D system with $H=V\oplus\openone$ where $V$
    decays with a power law $|\ld|^{-\nu}$, $\nu\ge2$, 
    the ground state correlations decay as $\Theta(|\ld|^{-\nu})$.
\end{itemize}
\end{theorem}
\end{minipage}

\begin{proof}
In all cases, we have to find the scaling of the ground state $\gamma$
which is the product $\gamma=(\hat\E^{-1}\oplus\hat\E^{-1})\sigma
H\sigma^T$, Eq.~(\ref{eq:basics:tinv-E0-and-gamma_from_Ehat}).
Part \textit{(i)} follows directly from Lemma~\ref{lemma:finite-H-exp-E},
as multiplying with a finite-range $\sigma H\sigma^T$ doesn't change the
exponential decay, while \textit{(ii)} follows from
Lemma~\ref{lemma:rapid-H-rapid-E}, the $o(\|n\|^{-\infty})$ decay of
$\sigma H\sigma^T$, and Lemma~\ref{lemma:poly-convolution}.
To show \textit{(iii)}, note that for $H=V\oplus\openone$, the ground
state is $\gamma=V^{-1/2}\oplus V^{1/2}$, and from
Lemma~\ref{lemma:poly-int-poly-decay}, $O(n^{-\nu})$ follows. 
For $\hat V^{-1/2}$, Lemma~\ref{lemma:poly-int-poly-decay} also includes
that the bound is exact, while for $\hat V^{1/2}$, it can be shown by
transferring the proof of the lemma one-to-one. 
\hspace*{\fill}\qed
\end{proof}

Note that a simple converse of Theorem~\ref{theorem:noncrit-decay}
always holds: for each translationally invariant pure state CM
$\gamma$, there exists a Hamiltonian $H$ with the same asymptotic
behavior as $\gamma$ such that $\gamma$ is the ground state of
$H$. This can be trivially seen by choosing
$H=\sigma\gamma\sigma^T$.

\section{Correlation length and gap
    \label{sec:corrlength-gap}}

In this section, we consider one-dimensional chains with local
gapped Hamiltonians. We compute the correlation length for these
systems and use this result to derive a relation between
correlation length and gap.

\begin{theorem}
\label{theorem:noncrit-1D-corrlength} Consider a non-critical 1D
chain with a local Hamiltonian. Define the complex extension of
the spectral function
$\E(\phi)=\Big[\sum_{\ld=0}^{L}c_\ld\cos(\ld\phi)\Big]^{1/2}$ in
Eq.~(\ref{eq:spectralfunction}) as
$$g(z):=\sum_{\ld=0}^{L}c_\ld\frac{z^\ld+z^{-\ld}}{2}\ ,$$
such that $g(e^{i\phi})=\E^2(\phi)
    \stackrel{(\ref{eq:basics:tinv-Ehat-definition})}{=}
    \hat H_Q(\phi)\hat H_P(\phi)-\hat H_{QP}^2(\phi)$ and let $\tilde z$ be zero of $g$
with the largest magnitude smaller than one. Then, the correlation length
$$\xi=-\frac{1}{\log|\tilde z|}$$ determines the asymptotic scaling of the
correlations which is given by
\begin{itemize}
\item $O^*(e^{-n/\xi}/\sqrt{n})$, if $\tilde z$ is a zero of
    order one, 
\item $O^*(e^{-n/\xi})$, if $\tilde z$ is a zero of
    even order, 
\item $o(e^{-n/(\xi+\eps)})$ for all $\eps>0$,
    if $\tilde z$ is a zero of odd order larger than one.
\end{itemize}
\end{theorem}

For the nearest neighbor interaction Hamiltonian ${\cal H}_\kappa$
from Eq.~(\ref{eq:basics:kleingordon}) one has for instance
$\E(\phi)=\sqrt{1-\kappa\cos(\phi)}$, so that $g$ has simple zeros
at $z_0=(1\pm\sqrt{1-\kappa^2})/{\kappa}$. Therefore $\tilde
z=({1-\sqrt{1-\kappa^2}})/{\kappa}$, and the correlations decay as
$\Theta(e^{-n/\xi}/\sqrt{n})$ where $\xi=-1/\log|\tilde z|$.

\begin{proof}
For local Hamiltonians, the correlations decay as the
matrix elements of $\hat\E^{-1}$
[Eq.~(\ref{eq:basics:tinv-E0-and-gamma_from_Ehat})].
By Fourier transforming (\ref{eq:basics:tinv-Ehat-definition}),
$\E(\phi)=\sqrt{g(e^{i\phi)}}$,  with
$g(e^{i\phi})=\hat H_Q(\phi)\hat H_P(\phi)-\hat H_{QP}^2(\phi)
=\sum_{\ld=0}^{L}c_\ld\cos(\ld\phi)$
an even trigonometric polynomial
(we assume $c_L\ne0$ without loss of generality),
 and $\min(g(e^{i\phi}))=\Delta^2$.
We have to compute
\begin{eqnarray}
(\hat\E^{-1})_n&=& \frac{1}{2\pi}
    \int_0^{2\pi}\frac{1}{\E(\phi)}e^{i\ld\phi}\dd\phi \;=\;\frac{1}{2\pi i}
    \int_{\mc S^1}\frac{z^{n-1}}{\sqrt{g(z)}}\:\mathrm{d}z\ ,
\label{eq:noncrit:corrlenth:unitcircle-int}
\end{eqnarray}
where $\mc S^1$ is the unit circle.
The function $g(z)$ has a pole of order $L$ at zero and $2L$ zeros
altogether. Since $\min(g(\phi))=\Delta^2>0$, $g$ has no zeros on the unit
circle. As $g(z)=g(1/z)$, the zeros come in pairs, and $L$ of them are
inside the unit circle.  Also, the conjugate of a zero is a zero as well.
From each zero with odd multiplicity emerges a branch cut of $\sqrt{g(z)}$.  We
arrange all the branch cuts inside the unit circle such that they go
straight to the middle where they annihilate with another cut. In case $L$
is odd, the last cut is annihilated by the singularity of $\sqrt{g(z)}$
at $0$. If two zeros lie on a line, one cut curves slightly. A sample
arrangement is shown in Fig.~\ref{fig:slits}.

\begin{figure}
\sidecaption
\hspace*{2em}\includegraphics[width=4cm]{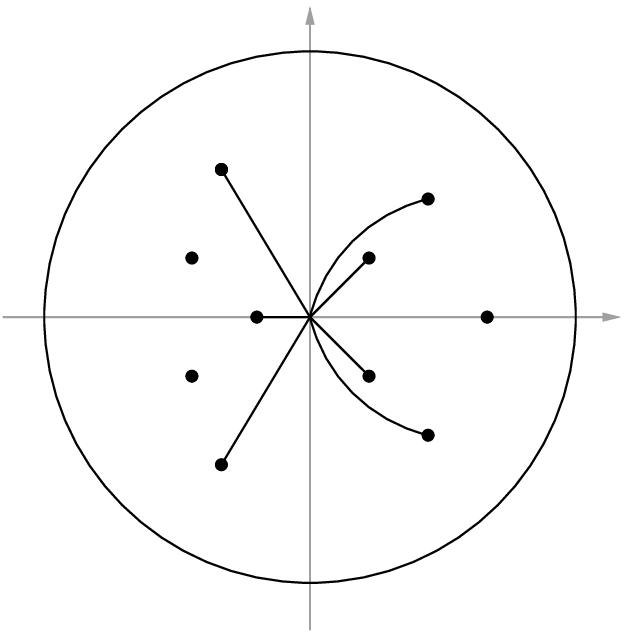}\hspace*{2em}
\caption{\label{fig:slits} Sample  arrangement of branch cuts 
and poles of $\sqrt{g}$ inside the unit circle.
From each odd order zero of $g$, a branch cut emerges. All
cuts go to $0$ where they cancel with another cut. In case their
number is odd, there is an additional branch point at $0$
cancelling the last cut. In case two zeros are on a line to the
origin, the cuts are chosen curved.  The integral of $\sqrt g$
around the unit circle is equal to the integral around the cuts,
plus integrals around the residues which originate from the even order
zeros of $g$.}
\end{figure}

Following Cauchy's theorem, the integral can be decomposed into
integrals along the different branch cuts and around the residues
of $1/\sqrt{g}$, and one has
to estimate the contributions from the different types of zeros
of $g$. The simplest case is given by zeros $z_0$ with even multiplicity
$2m$.  In that case, define $h(z):=g(z)/(z-z_0)^{2m}$ which has no zero
around $z_0$.  The contribution from $z_0$ to the correlations is then
given by the residue at $z_0$ and is
$$
\frac{1}{(m-1)!}\left.\frac{\dd^{m-1}}{\dd z^{m-1}}
    \left(\frac{z^{\ld-1}}{\sqrt{h(z)}}\right)
    \right|_{z=z_0}\propto z_0^{\ld-(m-1)}
$$
for $n-(m-1)>0$, i.e., it scales as $|z_0|^\ld$. Note that for
$z_0\not\in\mathbb R$, the imaginary parts originating from $z_0$ and its
conjugate
$\bar z_0$ exactly cancel out,
but the scaling is still given by $|z_0|^\ld=e^{\ld\log|z_0|}$, i.e.,
$\xi=-1/\log|z_0|$ is the corresponding correlation length.

If $z_0$ is a simple zero of $g(z)$, we have to integrate around
the branch cut. Assume first that the cut goes to zero in a
straight line, and consider a contour with distance $\eps$ to the
slit. Both the contribution from the $\eps$ region around zero and
the $\eps$ semicircle at $z_0$ vanish  as $\eps\rightarrow0$, 
and the total integral is
therefore given by twice the integral along the cut,
$$
\frac{1}{\pi i}\int_0^{z_0}\frac{z^{n-1}}{\sqrt{z-z_0}\sqrt{h(z)}}\dd z\ ,
$$
where again $h(z)=g(z)/(z-z_0)$. Intuitively, for growing $n$ the part of
the integral
close to $z_0$ becomes more and more dominating, i.e., the integral is
well approximated by the modified integral where $h(z)$ has been replaced
by $h(z_0)$. After rotating it onto the real axis, this integral---up to a
phase---reads
\begin{equation}
\frac{1}{\pi\sqrt{|h(z_0)|}}
\int_0^{|z_0|}\!\!\!\!\frac{r^{n-1}}{\sqrt{|z_0|-r}}\dd r=
\frac{|z_0|^{n-1/2}\Gamma(n)}{\sqrt{\pi|h(z_0)|}\;\Gamma(n+\tfrac12)}
\label{eq:noncrit:corrlenth:singlezero-gammabound}
\end{equation}
which for large $n$ is
\begin{equation}
\frac{1}{\sqrt{\pi|z_0h(z_0)|}}\
\frac{|z_0|^n}{\sqrt{n}}+O\left(\frac{|z_0|^n}{n^{3/2}}\right)\ .
\label{eq:noncrit:corrlenth:singlezero-sqrtbound}
\end{equation}
In order to justify the approximation $h(z)\leadsto h(z_0)$, consider the
difference of the two respective integrals. It is bounded by
$$
\left|
\int_0^{z_0}\frac{|z|^{n-1}}{\sqrt{|z-z_0|}}
    \smash{\underbrace%
    {\left|\frac{1}{\sqrt{h(z)}}-\frac{1}{\sqrt{h(z_0)}}\right|}_%
    {(*)}}
    \dd z
\right|\ .\rule[-3em]{0cm}{3em}
$$
On $[z_0/2,z_0]$, $h(z)$ is analytic and has no zeros, thus,
$|h(z)^{-1/2}-h(z_0)^{-1/2}|<C|z-z_0|$, where $C$ is the maximum of the
derivative of $h(z)^{-1/2}$ on $[z_0/2,z_0]$. On $[0,z_0/2]$,
the same bound is obtained by choosing $C$ the supremum of
$|h(z)^{-1/2}-h(z_0)^{-1/2}|/|z_0/2|$ on $[0,z_0/2]$.
Together, $(*)\le C|z-z_0|$, and the above integral is bounded by
$$
C\int_0^{|z_0|}r^{\ld-1}\sqrt{|z_0|-r}\,\dd r=
    C\frac{\sqrt\pi|z_0|^{\ld+1/2}\Gamma(\ld)}{2\Gamma(\ld+\tfrac32)}
    =O\left(\frac{|z_0|^n}{n^{3/2}}\right)\ ,
$$
i.e., it vanishes by $1/n$ faster than the asymptotics derived in
Eq.~(\ref{eq:noncrit:corrlenth:singlezero-sqrtbound}), which
justifies fixing $h(z)$ at $h(z_0)$.

From Eq.~(\ref{eq:noncrit:corrlenth:singlezero-sqrtbound}),
it follows that the scaling is
$e^{-\ld/\xi}/\sqrt{\ld}$, where the correlation length
is again $\xi=-1/\log|z_0|$.
The same scaling behavior can be shown to hold for appropriately
chosen curved branch cuts from $z_0$ to $0$ by relating the curved to a
straight integral.

The situation gets more complicated if zeros of odd order $>1$ appear.
In order to get an estimate which holds in all scenarios, we apply
Cauchy's theorem to contract the unit circle in the integration
(\ref{eq:noncrit:corrlenth:unitcircle-int}) to a circle of radius
$r>|z_0|$, where $z_0$ is the largest zero inside the unit
circle. Then, the integrand can be bounded by $C_r r^{\ld-1}$ (where
$C_r<\infty$ is the supremum of $1/\sqrt{g}$ on the circle), and this
gives a bound $2\pi C_r r^{\ld-1}$ for the integral.
This holds for all $r>|z_0|$, i.e., the correlation decay faster
than $e^{\ld \log r}$ for all $r>|z_0|$. This does not imply that the
correlations decay as $e^{\ld \log |z_0|}$, but it is still reasonable to
define $-1/\log|z_0|$ as the correlation length.
\hspace*{\fill}\qed
\end{proof}

\begin{theorem}\label{Thm:gap}
Consider a 1D chain together with a family of Hamiltonians
$H(\Delta)$ with gap $\Delta>0$, where $H(\Delta)$ is continuous
for $\Delta\rightarrow0$ in the sense that all entries of $H$
converge. Then, the ground state correlations scale exponentially,
and for sufficiently small $\Delta$ the correlation length is
$$
\xi\simeq\frac{1}{\sqrt{\Delta m^*}} \ .
$$
Here, $m^*=\left(\left.\frac{\mathrm d^2 \E(\phi)}
    {\mathrm d\phi^2}\right|_{\phi=\phi_\Delta}\right)^{-1}$
is the effective mass at the band gap.
\end{theorem}

For the discretized Klein-Gordon field
(\ref{eq:basics:kleingordon}), for example, we have
$\Delta=\sqrt{1-|\kappa|}$, $m^*=2\sqrt{1-|\kappa|}/|\kappa|$, and
for small $\Delta$ (corresponding to $|\kappa|$ close to 1), one obtains
$\xi\simeq\sqrt{|\kappa|/2(1-|\kappa|)}\simeq1/\sqrt{2}\Delta$.
Hence, the $\xi\propto 1/\Delta$ law holds if the coupling is
increased relative to the on-site energy (in which case
$m^*\propto\Delta$).

More generally, if we expand the spectral function
[Eq.~(\ref{eq:spectralfunction})] around the band gap we are
generically\footnote{This makes the natural assumption that the
minimum under the square root is quadratic. In fact, if it is of
higher order, then $m^*=\infty$ and thus $\xi=0$, which is
consistent with the findings of the following section. An example
of such a behavior is given by so called `quadratic interactions'
\cite{CEPD05} for which $H=V\oplus\openone$, where $V$ is the
square of a banded matrix.} led to the dispersion relation
$\E(k)\simeq\sqrt{\Delta^2+v^2 k^2}$ ($k\equiv\phi$). By the
definition of the effective mass and Theorem~\ref{Thm:gap} this
leads exactly to the folk theorem \be \xi\simeq\frac{v}\Delta\; .
\ee

\begin{proof}
According to Theorem~\ref{theorem:noncrit-1D-corrlength},
what remains to be done is to determine the position of
the largest zero $\tilde z$ of $g$ in the unit circle.
Due to the restriction on $H(\Delta)$, the coefficients of the polynomial
$g(z)z^L$ and thus  also the zeros of $g$ continuously depend on $\Delta$,
i.e., for sufficiently small $\Delta$, the zero closest to the unit circle
is the one closest to the gap.
In order to determine the position of this zero, we will expand
$g$ around the gap. We only discuss the generic case where the gap
appears only for one angle $\phi_0$, $g(\phi_0)=\Delta$. In the
case of multiple occurrences of the gap in the spectrum, one will
pick the gap which gives the zero closest to the unit circle,
i.e., the largest correlation length. Furthermore, we assume
$\phi_0=0$ without loss of generality. Otherwise, one considers
$g(ze^{-i\phi_0})$ instead of $g(z)$, which on the unit circle
coincides with the (rotated) spectrum.

The knowledge on $g=:u+iv$ (with $u,v:\mathbb{C}\rightarrow\mathbb{R}$)
which will be used in the proof is
\begin{equation}
\begin{array}{r@{\;=\;}l@{\qquad}r@{\;=\;}l}
u(1)&\Delta^2 & v(1)&0\\
u_\phi(1)&0& v_\phi(1)&0\\
u_{\phi\phi}(1)&2\Delta/m^*>0& v_{\phi\phi}(1)&0
\end{array}
\label{eq:corrlength:knowledge-at-1}
\end{equation}
where the subscripts denote the partial derivative with respect to the
respective subscript (in Euclidean coordinates $z\equiv x+iy$, in
polar coordinates $z\equiv re^{i\phi}$). Note that $z=1$ is the point
where the gap appears, and that $g(e^{i\phi})=\E(\phi)^2$ is real. Therefore,
the derivatives of the imaginary part $v$ along the circle vanish, while
the derivatives of the real part $u$ are found to be
$u(1)=\E(0)^2=\Delta^2$, $u_\phi(1)=2\E(0)\E'(0)=0$, and
$u_{\phi\phi}(0)=2\E'(0)^2+2\E(0)\E''(0)=2\Delta/m^*$, where
$m^*=1/\E''(\phi)$ is the effective mass at the band gap.

We need to exploit the relation between Euclidean and polar coordinates,
$$
\begin{array}{l@{\ ;\quad}l}
g_x(1)=g_r(1)& g_y(1)=g_\phi(1)\\
g_{xx}(1)=g_{rr}(1)&g_{yy}(1)=g_{\phi\phi}(1)+g_r(1)\\
\end{array}
$$
and the Cauchy-Riemann equations $ u_x=v_y$, $u_y=-v_x$, and
$g_{xx}+g_{yy}=0$, which together with the information
(\ref{eq:corrlength:knowledge-at-1}) lead to
\begin{eqnarray*}
&u(1)=\Delta^2\ ;\ v(1)=0\ ; \\
&u_x(1)=u_y(1)=v_x(1)=v_y(1)=0\ ;\\
&u_{xx}(1)=-2\Delta/m^*\ ;\ u_{yy}(1)=2\Delta/m^*\ ;\\
&v_{xx}(1)=0\ ;\ v_{yy}(1)=0\ .
\end{eqnarray*}
Note that it is not possible to derive information about the mixed second
derivates using only the information (\ref{eq:corrlength:knowledge-at-1}).
However, as long as $v_{xy}$ does not vanish at $1$, $v$ will only
stay zero in direction of $x$ or $y$, but not diagonally. Since
$\Delta^2>0$ and $2\Delta/m^*>0$, the closest zero is---to second
order---approximately located along the $x$ axis. By intersecting with the
parabola $\Delta^2-\frac{\Delta}{m^*}(x-1)^2$,
one finds that the zero is located at $x_0\approx1-\sqrt{\Delta m^*}$. For
small $\Delta$, the correlations thus decay with correlation length
$\xi\approx-1/\log(1-\sqrt{\Delta m^*})\approx1/\sqrt{\Delta m^*}$.
\hspace*{\fill}\qed
\end{proof}

\section{Critical systems
    \label{sec:critical}}

In the following, we discuss critical systems, i.e., systems
without an energy gap, $\Delta=0$.\footnote{Note that there are
different meanings of the notion criticality referring either to a
vanishing energy gap or to an algebraic decay of correlations. In
this section we discuss in which cases these two properties are
equivalent.} In that case, the Hamiltonian will get singular and
some entries of the ground state covariance matrix will diverge,
which leads to difficulties and ambiguities in the description of
the asymptotic behavior of correlations.  We will therefore
restrict to Hamiltonians of the type
$$
H=V\oplus\openone\ ,
$$
for which the ground state CM is $\gamma=V^{-1/2}\oplus V^{1/2}$.
While the $Q$ part diverges, the entries of the $P$-block stay
finite. Following Thm.~1(b) the extension to interactions of the
form $H=H_Q\oplus H_P$ is straight forward.

In order to compute the correlations we have to determine the
asymptotics of $V^{1/2}$, i.e.,
$$
(V^{1/2})_n=\frac{1}{(2\pi)^d}\int_{\mc T^d}
    \sqrt{\hat V(\novec\phi)}e^{i\novec \ld\novec\phi}\dd\novec\phi\ .
$$
We will restrict to the cases in which the excitation spectrum
$\E=\sqrt{\hat V}$ has only a finite number of zeros, i.e.,
finitely many points of criticality.  In addition, we will also
consider the special case in which the Hamiltonian exhibits a
tensor product structure.

We proceed as follows. First, we consider one-dimensional critical
chains and show that the correlations decay typically as
$O(\ld^{-2})$ and characterize those special cases where the
correlations decay more rapidly. The practically important case of
exactly cubic decaying interactions will be investigated in
greater detail. Depending on the sign of the interaction this case
will lead to a logarithmic deviation from the $n^{-2}$ behavior.
Then, we turn to higher dimensional systems and show that
generically the correlations decay as $\ld^{-(d+1)} \log\ld$, where
$d$ is the spatial dimension of the lattice.

\subsection{One dimension}

First, we prove a Lemma which shows that although taking the
square root of a smooth function destroys its differentiability,
the derivatives will stay bounded.

\begin{lemma}
\label{lemma:bounded-derivatives-of-sqrt-1d}
Let $f\in\mathscr C^m([-1;1])$,
$f(x)\ge0$ with the only zero at $x=0$, and
let $2\nu\le m$ be the order of the minimum at $x=0$,
i.e., $f^{(k)}(0)=0\ \forall k<2\nu$, $f^{(2\nu)}(0)>0$.

Define $g(x):=\sqrt{f(x)}$. Then, the following holds:
\begin{itemize}
\item For odd $\nu$, $g\in\mathscr C^{\nu -1}([-1;1])$,
and $g\in\C^{m-\nu }([-1;0])$, $g\in\C^{m-\nu }([0;1])$, i.e.,
the first $m-\nu $ derivatives (for $x\ne0$) are bounded.
\item For even $\nu $, $g\in\mathscr C^{m-\nu }([-1;1])$.
\end{itemize}
\end{lemma}

\begin{proof}
Using the Taylor expansion
$f(x)=\sum_{k=2\nu}^{m}c_kx^{k}+\rho(x)$,
$\rho^{(k)}(x)=o(x^{m-k})$ for $k\le m$, we express $g$ as
$g(x)=(\sgn\, x)^\nu x^\nu r(x)$ with
\[
    r(x)= \sqrt{\sum_{k=2\nu}^{m}c_kx^{k-2\nu}+\frac{\rho(x)}{x^{2\nu}}}\ ,
\]
where we used that $(\sgn\,x)^\nu x^\nu =\sqrt{x^{2\nu}}$.
Let us now consider the derivatives of $r(x)$. While the sum leads
to a $O(1)$ contribution, the $k$'th derivative of the remainder
behaves as $o(1)/x^{2\nu -m+k}$. Together, this leads to
\[ \begin{array}{l@{\qquad}l}
r^{(k)}(x)=O(1)               &  2\nu -m+k\le 0\ ,\\
r^{(k)}(x)=o(1)/x^{2\nu -m+k}    &  2\nu -m+k\ge 1\ .
\end{array} \]
Now consider the $k$'th derivative of $g(x)$ for $x\ne0$,
\[
g^{(k)}(x)=(\mathrm{sgn}\, x)^\nu
    \sum_{l=0}^k\underbrace{{k\choose l}
    \left[\frac{\dd^l}{\dd x^l}x^\nu \right] r^{(k-l)}(x)}_{s_l}\ .
\]
Assume first $k\le \nu $. Then, $s_l\propto O(1)x^{\nu -l}$ for $2\nu
-m+k-l\le0$, and $s_l\propto o(1)x^{m-\nu -k}$ for $2\nu
-m+k-l\ge1$, and as $m\ge 2\nu $, it follows that $g^{(k)}=O(x)$
for $k<\nu $, which cancels the discontinuity originating from
$\sgn\, x$. For $k=\nu $, on the contrary, $s_k=O(1)$, and $\sgn\,
x$ introduces a discontinuity on $g^{(k)}$, yet, it remains
bounded and piecewise differentiable on $[-1;0]$ and $[0;1]$. The
first non-bounded $s_l$ is found as soon as $m-\nu -k=-1$, and
$g\in\C^{m-\nu }([0;1])$ directly follows.

This also implies that for $m-\nu-k\ge0$, ${g(x)}/{(\sgn\,
x)^\nu}\in \C^{m-\nu}([-1;1])$, i.e., the discontinuity is only
due to $(\sgn\, x)^\nu$. Since, however, this is only
discontinuous for odd $\nu$, it follows that
$g\in\C^{m-\nu}([-1;1])$ if $\nu$ even.
\hspace*{\fill}\qed
\end{proof}

\begin{theorem}
\label{theorem:crit:1D}
Consider a one-dimensional critical chain with
Hamiltonian $H=V\oplus\openone$, where $V_n=O(n^{-\alpha})$, $\alpha>4$ 
and where $\hat V$ has a finite number of critical points which are
all quadratic minima of $\hat V$.  Then, $(\gamma_P)_n=O^*(n^{-2})$. 
For $V_n\propto n^{-\alpha}$, $\alpha>3$ it even follows
that $(\gamma_P)=\Theta(n^{-2})$.
\end{theorem}

Note that for $V_n\propto n^{-\alpha}$, the extrema of $\hat V$
are always quadratic.

\begin{proof}
We want to estimate
\begin{equation}
(V^{1/2})_n=\frac{1}{2\pi}
    \int_{\mc S^1}g(\phi)e^{i\ld\phi}\mathrm d\phi\ ,
\label{eq:critical_int}
\end{equation}
where $g=\hat V^{1/2}$. Under both assumptions, $\hat V\in\C^2(\mc
S^1)$, and all critical points are minima of order $2$.  It follows from
Lemma~\ref{lemma:bounded-derivatives-of-sqrt-1d} that $g$ is
continuous with bounded derivative. Therefore, we can
integrate by parts, the bracket vanishes, and we obtain
$$
(V^{1/2})_n=-\frac{1}{2\pi i\ld}\int_0^{2\pi}
g'(\phi)e^{i\ld\phi}\mathrm d\phi\ .
$$

Now, split $\mc S^1$  at the zeros of $g$ into closed intervals
$\mc I_j$, $\bigcup_j \mc I_j=\mc S^1$, and rewrite the above
integral as a sum of integrals over $\mc I_j$.  As $g'\in\C(\mc
I_j)$ (and differentiable on the inner of $\mc I_j$), one can once
more integrate by parts which yields
\begin{equation}
\label{eq:crit:1D:last-integral}
(V^{1/2})_n=-\frac{1}{2\pi(i\ld)^2}\sum_{j}\Bigg(
    \left[g'(\phi)e^{i\ld\phi}\right]_{\mc I_j}-
    \int_{\mc I_j} g''(\phi)e^{i\ld\phi}\mathrm d\phi
    \Bigg)\ .
\end{equation}
Neither of the terms will vanish, but since $g'\in\C(\mc I_j)$,
the bracket is bounded. In case $V_n\in O(n^{-\alpha})$, $\alpha>4$,
we have $\hat V\in\C^3(\mc S^1)$, therefore $g''$ is bounded
(Lemma~\ref{lemma:bounded-derivatives-of-sqrt-1d}), and the integrals
vanish as $o(1)$. Unless the contributions of the brackets for the
different $\mc I_j$ cancel out, the $n^{-2}$ bound is tight,
$(V^{1/2})_n=O^*(n^{-2})$. The tightness of the bound is also illustrated
by the example which follows the proof.

For the case of an exactly polynomial decay, we additionally have to show
that $g''$ is absolutely integrable for $3<\alpha\le4$. Then, the exactness of
the bound holds because the bracket in
Eq.~(\ref{eq:crit:1D:last-integral}) does not
oscillate (the critical point is either at $\phi=0$
or at $\phi=\pi$), and because the integral is $o(1)$ for $g''\in\mc
L^1(\mc S^1)$.  In case the critical point is at $\phi=\pi$, 
the latter holds since $\hat V\in\C^\infty((0;2\pi))$ implies that  $g''$
is bounded at $\pi$, and $\hat V\in\C^2(\mc S^1)$ that
$g\in\C^2((-\pi,\pi))$, which together proves that $g''$ is bounded on
$\mc S^1$. 

In case the critical point is at $\phi=0$, the situation is more involved
(and for $\alpha=3$, a logarithmic correction appears,
cf.~Theorem~\ref{theorem:crit:iontrap-logcorrection}). 
Since $\hat V^{(3)}(\phi)=
-\mathrm{Im}\big[\mathrm{Li}_{\alpha-3}(e^{i\phi})\big]=
O(\phi^{\alpha-4})$, we have 
$$
\hat V''(\phi)=\hat V''(0)+O(\phi^{\alpha-3}),\;
\hat V'(\phi)=\hat V''(0)\phi+O(\phi^{\alpha-2}),\;
\hat V(\phi)=\tfrac12\hat V''(0)\phi^2+O(\phi^{\alpha-1}).
$$
With this information,
$$g''(\phi)=
    \frac{2\hat V(\phi)\hat V''(\phi)-\hat V'(\phi)^2}{4V(\phi)^{3/2}}
    =O(\phi^{\alpha-4})\ ,
$$
which indeed proves that $g''\in\mc L^1(\mc S^1)$, and thus
$(V^{1/2})_n=\Theta(n^{-2})$.
\hspace*{\fill}\qed
\end{proof}

As an example, consider again the discretized Klein-Gordon field
of Eq.~(\ref{eq:basics:kleingordon}) which is critical for
$\kappa=\pm1$, corresponding to $\hat V(\phi)=1\mp\cos\phi$. The
Fourier integral is solvable and yields
$(\gamma_P)_n=-\frac{2\sqrt{2}}{\pi}
    \frac{(\sgn\,\kappa)^n}{4n^2-1}=\Theta(n^{-2})$.

\vspace{2ex}

\textbf{Generalizations of Theorem~\ref{theorem:crit:1D}.} Using
Lemma~\ref{lemma:bounded-derivatives-of-sqrt-1d}, several
generalizations for the 1D critical case can be found. In the following,
we mention some of them. In all cases $H=V\oplus\openone$ is critical.\\
\textit{Critical points of even order.}---If $V_n=o(n^{-\infty})$ and the
critical points are minima of order $2\nu$, $\nu$ even, the correlations
decay as $(\gamma_P)_n=o(n^{-\infty})$. This is the case, e.g., if 
$V=X^2$ with $X$ itself rapidly decaying. 
\\
\textit{Critical points of higher order.}---If $\hat V$ has critical
points of order at least $2\nu$, $\nu$ odd, and $V_n=O(n^{-\alpha})$,
$\alpha>2\nu+2$, then $(\gamma_P)_n=O(n^{-(\nu+1)})$.\\
\textit{Minima of different orders.}---If $\hat V$ has minima of different
orders $2\nu_i$, in general the minimum with the lowest odd
$\nu_i\equiv\nu_1$ will determine the asymptotics,
$(\gamma_P)_n=O(n^{-(\nu_1+1)})$. As $\hat V\in\C^{(2\max\{\nu_i\})}(\mc
S^1)$ is required anyway, the piecewise differentiability of $\hat
V^{1/2}$ is guaranteed.\\
\textit{Weaker requirements on $V$.}---It is possible to ease the
requirements imposed on $V$ in Theorem~\ref{theorem:crit:1D} to 
$V_n=O(n^{-\alpha})$, $\alpha>3$ or $\hat V\in\C^2(\mc S^1)$,
respectively. The price one has to pay is that one gets an additional
$\log$ correction as in the multidimensional critical case,
Theorem~\ref{theorem:critical-multidim}. The method to bound $g''$ is the
same which is used there to derive (\ref{eq:crit:deriv-bound}).

\vspace{1.5ex}

The above proof does not cover the case of the relevant
$1/n^3$ interaction, which for instance appears for the motional degrees
of freedom of trapped ions. In the following, we separately discuss this
case.  It will turn out that the scaling will depend on the sign of the
coupling: while a positive sign (corresponding to the radial degrees of
freedom) again gives a $\Theta(\tfrac{1}{n^2})$ scaling as before, for the
negative sign (corresponding to the axial degree of freedom) one gets
$\Theta\big(\tfrac{\sqrt{\log n}}{n^2}\big)$.

\begin{theorem}
Consider a critical 1D chain with a $1/n^3$ coupling with positive
sign, i.e., $H=V\oplus\openone$, $V_n=c/n^3$, $V_0=3c\zeta(3)/2$,
$c>0$, with $\zeta$ the Riemann zeta function. Then, the ground
state correlations scale as
$(\gamma_{P})_n=\Theta(\tfrac{1}{n^2})$.
\end{theorem}

\begin{proof}
We take w.l.o.g.\ $c=1/2$. For this sign of the coupling, the critical
point is at $\pi$, $\hat V(\pi)=0$.
From the proof of Lemma~\ref{lemma:poly-int-poly-decay}, we know
that $\hat V\in\C^1(\mc S^1)$, $\hat V\in\C^\infty((0;2\pi))$, and
that $\hat V''(\phi)=\log(2\sin(\phi/2))$ on $(0;2\pi)$. 
With $g:=\hat V^{1/2}$,
it follows from
Lemma~\ref{lemma:bounded-derivatives-of-sqrt-1d} 
that $g\in\C(\mc S^1)$, 
$g\in\C^1([-\pi;\pi])$, 
and $g\in\C^\infty((0;\pi])$, $g\in\C^\infty([-\pi;0))$.
This means that all derivatives $g^{(k)}$, $k\ge1$ can exhibit jumps at
the critical point $\pi$ but they all remain bounded. In contrast, around
$\phi=0$, $g'$ is continuous but $g''$ has a $\log$ divergence.

Thus, the
Fourier integral
$$
(V^{1/2})_n=\frac{1}{2\pi}\int_{\mc S^1}
    g(\phi)e^{in\phi}\dd \phi
$$
can be split at $0$ and $\pi$, and then integrated by parts
twice. The brackets of the first integration cancel out due to
continuity of $g$, and one remains with
$$
(V^{1/2})_n= \frac{1}{\pi(in)^2}\left(
    \left[g'(\phi)\cos(n\phi)\right]_0^\pi+
    \int_{0}^{\pi}g''(\phi)\cos(n\phi)\dd\phi
    \right)\ ,
$$
where we used the symmetry of $g$. One finds
$[g'(\phi)\cos(n\phi)]_0^\pi=-\sqrt{\tfrac{\log 2}{2}} (-1)^n$, 
and since $g''$ is integrable,
the integral is $o(1)$ due to the Riemann-Lebesgue lemma. 
Together, 
this proves $(\gamma_P)_n=\Theta(\tfrac{1}{n^2})$.  
\hspace*{\fill}\qed
\end{proof}

\begin{theorem}
\label{theorem:crit:iontrap-logcorrection}
Consider a critical 1D chain with a $1/n^3$ coupling with negative
sign, i.e., $H=V\oplus\openone$, $V_n=-c/n^3$, $V_0=2c\zeta(3)$,
$c>0$, with $\zeta$ the Riemann zeta function. Then, the ground
state correlations scale as
$(\gamma_P)_n=\Theta\big(\tfrac{\sqrt{\log n}}{n^2}\big)$.
\end{theorem}

\begin{proof}
Again, take w.l.o.g.\ $c=1/2$. For the negative sign of the interaction,
the critical point is at $\phi=0$. Since at this point $\hat V''$
diverges, Lemma~\ref{lemma:bounded-derivatives-of-sqrt-1d} cannot be
applied, and the situation gets more involved.

As in the previous proof, we use that $\hat V\in\C^1(\mc S^1)$,
$\hat V\in\C^\infty((0;2\pi))$, and thus $\hat V^{1/2}\in\C(\mc
S^1)$, $\hat V^{1/2}\in\C^\infty((0;2\pi))$. 
Further, $\hat V''(\phi)=-\log(2\sin(\phi/2))$ on
$(0;2\pi)$, cf.\ the proof of
Lemma~\ref{lemma:poly-int-poly-decay}, and with $\sin
x=x(1+O(x^2))$ we have $\hat V''(\phi)=-\log(\phi)+O(\phi^2)$ for
$\phi\rightarrow0$ (and similarly for $\phi\rightarrow2\pi$), and
therefore
\begin{equation}
\hat V'(\phi)=\phi(1-\log\phi)+O(\phi^3)\;,\quad
\hat V(\phi)=\tfrac14\phi^2(3-2\log\phi)+O(\phi^4)\ .
\label{eq:crit:1D:ionchain-Vhat-approx}
\end{equation}
As $\hat V^{1/2}\in\C(\mc S^1)$, we can integrate by parts one time,
\begin{equation}
(V^{1/2})_n=\frac{1}{2\pi}\int_{\mc S^1}
    \hat V^{1/2}(\phi)e^{in\phi}\dd \phi=
    \frac{1}{\pi n}\int_{0}^{\pi}
    g'(\phi)\sin(n\phi)\dd \phi
\label{eq:crit:1D:ionchain:after-first-partial}
\end{equation}
where we exploited the symmetry of $\hat V$, and with $g:=\hat V^{1/2}$.
Then, from (\ref{eq:crit:1D:ionchain-Vhat-approx}),
$$
g'(\phi)=\frac{1-\log\phi}{\sqrt{3-2\log\phi}}+
    O\bigg(\frac{\phi^2}{\sqrt{|\log\phi|}}\bigg),\
g''(\phi)=\frac{-2+\log\phi}{\phi(3-2\log\phi)^{3/2}}
    +O\bigg(\frac{\phi}{\sqrt{|\log\phi|}}\bigg),
$$
and after another round of approximation,
$$
g'(\phi)=\frac{\sqrt{|\log\phi|}}{\sqrt{2}}+
    O\bigg(\frac{1}{\sqrt{|\log\phi|}}\bigg),\
g''(\phi)=-\frac{1}{2^{3/2}}\frac{1}{\phi\sqrt{|\log\phi|}}+
    O\bigg(\frac{1}{\phi|\log\phi|^{3/2}}\bigg).
$$
This shows that the remainder $g'(\phi)-\sqrt{|\log\phi|/2}$ is continuous
with a absolutely integrable derivative, and by integration by parts
it follows that it only leads to a contribution $O(1/n)$ in
the integral (\ref{eq:crit:1D:ionchain:after-first-partial}). 
Thus, it remains to
investigate the asymptotics of the sine Fourier coefficients of
$h(\phi)=\sqrt{|\log\phi|}$. For convenience, we split the
integral (\ref{eq:crit:1D:ionchain:after-first-partial}) at $1$,
and $[1;\pi]$ only contributes with $O(1/n)$, as $h$ is continuous
with absolutely integrable derivative on $[1;\pi]$. On $[0;1]$, we
have to compute the asymptotics of
\begin{equation}
\label{eq:crit:1D:the-sqrt-log-integral}
\mc I=\int_0^1\sqrt{-\log\phi}\sin(n\phi)\dd \phi\ .
\end{equation}
Therefore, split the integral at $1/n$. The left integral can be
bounded directly, and the right after integration by parts [cf.\ 
the treatment of Eq.~(\ref{eq:crit:g-fourier-int})].  One gets
$$
\mc I\le
    \int_{0}^{1/n}\sqrt{-\log\phi}\;\dd\phi+
    \frac{\sqrt{\log n}}{n}+
    \frac1n\int_{1/n}^1\frac{1}{2\phi\sqrt{-\log\phi}}\dd\phi=
    O\left(\frac{\sqrt{\log n}}{n}\right)\ .
$$
In order to prove that this is also a lower bound for the asymptotics, it
suffices to show this for the integral
(\ref{eq:crit:1D:the-sqrt-log-integral}) as all other contributions vanish
more quickly. To this end, split the integral
(\ref{eq:crit:1D:the-sqrt-log-integral}) into single oscillations of the
sine, $J_k=[\tfrac{2\pi k}{n}, \tfrac{2\pi (k+1)}{n}]$, $k\ge0$. As
$\sqrt{-\log\phi}$ has negative slope on $(0;1)$, each of the $J_k$ gives
a positive contribution to $\mc I$, and thus we can truncate the integral
at $\tfrac12$,
\begin{equation}
\label{eq:crit:1D:sum-for-sqrt-log-lower-bound}
\mc I\ge\sum_{\frac{2\pi(k+1)}{n}\le\frac12}
    \int_{J_k}\sqrt{-\log\phi}\,\sin(n\phi)\,\dd\phi\ .
\end{equation}
On $[0;\tfrac12]$, $\sqrt{-\log\phi}$ has a positive curvature, and thus,
each of the integrals can be estimated by linearly approximating
$\sqrt{-\log\phi}$ at the middle of each $J_k$ but with the slope at
$\tfrac{2\pi (k+1)}{n}$, which gives
$$
\int_{J_k}\sqrt{-\log\phi}\,\sin(n\phi)\,\dd\phi\ge
    \frac{\pi}{n^2}\frac{1}{\tfrac{2\pi(k+1)}{n}
        \sqrt{-\log\left[\tfrac{2\pi(k+1)}{n}\right]}}\ .
$$
Now, we plug this into the sum
(\ref{eq:crit:1D:sum-for-sqrt-log-lower-bound}) and bound the sum by the
integral from $\tfrac{2\pi}{n}$ to $\tfrac12$ (the integrand in
monotonically decreasing), which indeed gives a lower bound
$\tfrac{1}{n}(\sqrt{\log\tfrac{n}{2\pi}}-\sqrt{\log 2})$ on $\mc I$ 
and thus proves the $\Theta(\sqrt{\log n}/n^2)$ scaling.
\hspace*{\fill}\qed
\end{proof}

\subsection{Higher dimensions}

For more than one dimension, the situation is more involved. First
of all, it is clear by taking many uncoupled copies of the
one-dimensional chain that there exist cases where the
correlations will decay as $\ld^{-2}$. However, these are very
special examples corresponding  to Hamiltonians with a tensor
product structure $H_{i_1i_2,j_1j_2}=H_{i_1,j_1}H'_{i_2,j_2}$. In contrast, we show that for generic systems the correlations in the
critical case decay as $O(\ld^{-(d+1)}\log\ld)$, where $d$ is the
dimension of the lattice. The requirement is again that the energy
spectrum $\E (\novec\phi)$ has only a finite number of zeroes,
i.e., finitely many critical points.

Note that the case of a Hamiltonian with a tensor product
structure can also be solved, as in that case $\hat V$ becomes a
product of terms depending on one $\phi_i$ each and thus the
integral factorizes. Interestingly, although the correlations
along the axes decay as $\ld^{-2}$, the correlations in a fixed
diagonal direction will decay as $\ld_1^{-2}\cdots
\ld_d^{-2}\propto \|\novec \ld\|^{-2d}$ and thus even faster than
in the following theorem. The $O\big(\|\novec
\ld\|^{-(d+1)}\log\|\ld\|\big)$ decay of the theorem holds
isotropically, i.e., independent of the direction of $\novec \ld$.

\begin{theorem}
\label{theorem:critical-multidim} Consider a $d$-dimensional
bosonic lattice with a critical Hamiltonian $H=V\oplus \openone$.
Then the $P$-correlations of the ground state decay as
$$
O\left(\|\novec \ld\|^{-(d+1)}\log \|\novec \ld\|\right)
$$
if the following holds: $\hat V\in\mathscr C^{d+1}$
[e.g., the correlations decay as $O(\|\ld\|^{-(2d+1+\eps)})$,
$\eps>0$], $\hat V$ has only a finite number of zeros which are
quadratic minima, i.e., the Hessian
$\left(\frac{\partial^2 \hat V(\novec\phi)}%
{\partial \phi_i\partial\phi_j}\right)_{ij}$ is positive definite
at all zeros.
\end{theorem}

\begin{proof}
We have to evaluate the asymptotic behavior of the
integral
$$
(\hat V^{1/2})_n=
    \frac{1}{(2\pi)^d}
    \int_{\mc T^d}\dd^d\novec\phi
    \sqrt{\hat V(\novec\phi)}\cos[\novec \ld\novec\phi]\ .
$$
Let us first briefly sketch the proof. We start by showing that it
suffices to analyze each critical point separately.  To this end, we show
that is is possible to smoothly cut out some environment of each critical
point which reproduces the asymptotic behavior. Then, we rotate the
coordinate system such that we always look at the correlations in a fixed
direction, and integrate by parts---which
surprisingly can be carried out as often as $\hat V$ is differentiable,
as all the brackets
vanish. Therefore, the information about the asymptotics is contained in
the remaining integral, and after a properly chosen number of partial
integrations, we will attempt to estimate this term.

Let now $\novec\zeta_i$, $i=1,\dots,I$ be the zeros of $\hat V$.
Clearly, these will be the only points which contribute to the
asymptotics as everywhere else $\sqrt{\hat V}$ is $\C^{d+1}$. In
order to separate the contributions coming from the different
$\novec\zeta_i$, we will make use of so-called
\emph{neutralizers}~\cite{BH86}. For our purposes, these are
functions $\mc N_{\novec \xi_0,r}\in \C^\infty(\mathbb
R^d\rightarrow[0;1])$ which satifsy
$$
\mc N_{\novec \xi_0,r}(\novec \xi)=\left\{\begin{array}{l@{\ :\ }l}
    1&\|\novec \xi-\novec \xi_0\|\le r/2\\
    0&\|\novec \xi-\novec \xi_0\|\ge r
    \end{array}\right.
$$
and are rotationally symmetric (cf.~\cite{BH86} for an explicit
construction). For each $\novec\zeta_i$, there exists an $r_i$
such that the balls $B_{r_i}(\novec\zeta_i)$ do not intersect. We
now define the functions
$$
f_i(\novec\phi):=\sqrt{\hat V(\novec\phi)}\
    \mc N_{\novec\zeta_i,r_i}(\novec\phi) \; ,\quad
\rho(\novec\phi):=\sqrt{\hat
V(\novec{\phi})}-\sum_{i=1}^If_i(\novec\phi)\; .
$$
Clearly, $\rho$ is $\C^{d+1}$, and so is each $f_i$ except at $\zeta_i$.
Furthermore, each $f_i$ is still the square root of a $\mathscr \C^{d+1}$
function.  By definition,
\begin{equation}
(\hat V^{-1/2})_n=
    \frac{1}{(2\pi)^d}\sum_{i=1}^I
    \int_{\mc T^d}\dd^d\phi f_i(\phi)\cos[n\phi]+
    \frac{1}{(2\pi)^d}\int_{\mc T^d}\dd^d\phi \rho(\phi)\cos[n\phi]\ ,
\label{eq:crit:corr-element-sum}
\end{equation}
i.e., it suffices to look at the asymptotics of each $f_i$ separately. The
contribution of $\rho$ is $O(\|\novec \ld\|^{-(d+1)})$ as can be
shown by successive integrations by parts just as for the non-critial
lattice (cf.\ the proof of Lemma~\ref{lemma:rapid-H-rapid-E}).

Let us now analyze the integrals
$$
I_i=\int_{B_{r_i}(\novec\zeta_i)}\dd^d\novec\phi
    f_i(\novec\phi)\cos[\novec \ld\novec\phi]\ .
$$
The integration range can be restriced to $B_{r_i}(\novec\zeta_i)$ as $f_i$
vanishes outside the ball.  By a rotation, this can be mapped to an
integral where $\novec \ld=(\|\ld\|,0,\dots,0)$, whereas $f_i$
is rotated to another function $\tilde f_i$ with the same properties,
$$
I_i=\int_{B_{r_i}(\zeta_i)}\dd^d\novec\phi\tilde
f_i(\novec\phi)\cos[\|\ld\|\phi_1]\ .
$$
Since the integrand is continuous and thus bounded, it is absolutely
integrable, and from Fubini's theorem, one finds
$$
I_i=\int\limits_%
    {\makebox[0cm][c]{\footnotesize$B_{r_i}(\novec{}\tilde\zeta_i)$}}
    \dd^{d-1} \novec{}\tilde\phi
    \underbrace{
    \int\limits_{\zeta_{i,1}-r_i}^{\zeta_{i,1}+r_i}\dd
    \phi_1\tilde f_i(\phi_1,\novec{}\tilde\phi)\cos[\|\ld\|\phi_1]}_%
    {J_i(\novec{}\tilde\phi)}\;,
$$
where we separated out the integration over the first component. The
vector $\novec{}\tilde\phi$ denotes the components $2\dots d$ of $\novec\phi$. The
extension of the integration range to a cylinder is possible as $\tilde
f_i$ vanishes outside $B_{r_i}(\novec\zeta_i)$.

Let us now require $\novec{}\tilde\phi\ne\novec{}\tilde\zeta_i$. This does not change the
integral since the excluded set is of measure zero, but it ensures that
$\tilde f_i$ is in $\C^{d+1}$. This allows us to integrate the
inner integral $J_i(\novec{}\tilde\phi)$ by parts up to $d+1$ times, and
each of the brackets
$$
\left[\tilde f_i^{(k)}(\phi_1,\novec{}\tilde\phi)\frac{1}{\|\ld\|^k}
    \cos(\|\ld\|\phi_1-k\pi/2)
    \right]_{\phi_1=\zeta_{i,1}-r_i}^{\zeta_{i,1}+r_i}
$$
appearing in the $k$'th integration step vanishes. Here, $\tilde
f_i^{(d)}(\phi_1,\novec{}\tilde\phi)= \partial^d\tilde
f_i(\phi_1,\novec{}\tilde\phi)/\partial\phi_1^d$ is the $d$'th
partial derivative with respect to the first argument. After
integrating by parts $d$ times, we obtain
\begin{equation}
I_i=\frac{1}{\|\ld\|^{d}}\int\limits_%
    {\makebox[1em][c]{\footnotesize$B_{r_i}(\novec{}\tilde\zeta_i)$}}
    \dd^{d-1} \novec{}\tilde\phi
    \int\limits_%
    {\makebox[1em][c]{\footnotesize$\zeta_{i,1}-r_i$}}^%
    {\makebox[1em][c]{\footnotesize$\zeta_{i,1}+r_i$}}
        \dd \phi_1
        \tilde f_i^{(d)}(\phi_1,\novec{}\tilde\phi)
        \cos[\|\ld\|\phi_1-d\pi/2]\ .
\label{eq:crit:int-after-partial}
\end{equation}

Now we proceed as follows: first, we show that the order of
integration can be interchanged, and second, we show that for the
function obtained after integrating $\tilde f_i^{(d)}$ over
$\novec{}\tilde\phi$, the Fourier coefficients vanish as
$\log(\|\ld\|)/\|\ld\|$.

The central issue for what follows is to find suitable bounds on
$|\tilde f_i^{(k)}|$.  Therefore, define $\tilde f_i^2=:h_i\in
\C^{d+1}$. By virtue of Taylor's theorem, and as $h_i(\novec\zeta_i)=0$ is
a minimum,
$$
h_i(\novec\phi)=
    {\textstyle\frac12}
    (\novec\phi-\novec\zeta_i)\cdot(\mathbf{D}^2h_i(\novec\zeta_i))
        (\novec\phi-\novec\zeta_i)+
    o(\|\novec\phi-\novec\zeta_i\|^2)
$$
with $\mathbf{D}^2$ the second derivative.  As the first term
is bounded by $\frac12\|\mathbf{D}^2h_i(\zeta_i)\|_\infty
\|\phi-\zeta_i\|^2$ and the second vanished faster than
$\|\novec\phi-\novec\zeta_i\|^2$, we can find $\eps_i>0$ and $C_1>0$ such that
\begin{equation}
\label{eq:crit:fcond-square-above}
|h_i(\novec\phi)|\le C_1\|\novec\phi-\novec\zeta_i\|^2
    \qquad\forall \|\novec\phi-\novec\zeta_i\|<\eps_i\ .
\end{equation}
By looking at the Tayor series of $h_i'\equiv\partial h_i/\partial \phi_1$
up to the first order we also find that there are $\eps_i>0$ and
$C_2>0$ such that
\begin{equation}
\label{eq:crit:fcond-lin-above}
|h_i'(\novec\phi)|\le C_2\|\novec\phi-\novec\zeta_i\|
    \qquad\forall \|\novec\phi-\novec\zeta_i\|<\eps_i\ .
\end{equation}
In addition to these upper bounds, we will also need a lower bound on
$|h_i|$. Again, by the Taylor expansion of $h_i$ around $\zeta_i$, we
find
$$
|h_i(\novec\phi)|\ge \lambda_{\min}\left[\mathbf{D}^2h_i(\novec\zeta_i)\right]-
    o(\|\novec\phi-\novec\zeta_i\|^2)\ ,
$$
and as all the zeros are quadratic minima, i.e.,
$\lambda_{\min}\left[\mathbf{D^2}h_i(\novec\zeta_i)\right]>0$, there exist
$\eps_i>0$, $C_3>0$ such that
\begin{equation}
\label{eq:crit:fcond-square-below}
|h_i(\novec\phi)|\ge C_3\|\novec\phi-\novec\zeta_i\|^2
    \qquad\forall \|\novec\phi-\novec\zeta_i\|<\eps_i\ .
\end{equation}
Clearly, $\eps_i$ can be chosen equal in
Eqs.~(\ref{eq:crit:fcond-square-above}-\ref{eq:crit:fcond-square-below}).
Note that the bounds can be chosen to be invariant under rotation of $h_i$
and thus of $\tilde f_i$. This holds in particular for the $\eps_i$ as the
remainders of Taylor series vanish uniformly.  Thus, the bound we will
obtain for the correlation function indeed only depends on $\|\ld\|$
and not on the direction of $n$.

Now, we use the conditions (\ref{eq:crit:fcond-square-above}-%
\ref{eq:crit:fcond-square-below}) to derive bounds on $|\tilde f_i^{(k)}|$.
Therefore, note that from $\tilde f_i\equiv\sqrt{h_i}$ it follows
that
$$
\tilde f_i^{(k)}=\frac{%
    \sum\limits_{{j_1+\dots+j_k=k}\atop{j_\nu=0,1,2,\dots}}
    c_{j_1\dotsc j_k}h_i^{(j_1)}\cdots h_i^{(j_k)}
    }{%
    h_i^{(2k-1)/2}
    }\ .
$$
One can easily check that for each term in the numerator, the
number $K_0$ of zeroth derivatives and the number $K_1$ of first
derivatives of $h_i$ satisfy $2K_0+K_1\ge k$. By bounding all
higher derivatives of $h_i$ from above by constants, we find that
the modulus of each summand in the numerator, and thus the modulus
of the numerator itself, can be bounded above by
$C'\|\novec\phi-\novec\zeta_i\|^k$  in the ball
$B_{\eps_i}(\zeta_i)$ with some $C'>0$.  On the other hand, it
follows directly from (\ref{eq:crit:fcond-square-below}) that the
modulus of the denominator is bounded below by
$C''\|\novec\phi-\novec\zeta_i\|^{2k-1}$, $C''>0$, such that in
total
\begin{equation}
|\tilde f_i^{(k)}(\novec\phi)|\le C\frac{1}{\|\novec\phi-\novec\zeta_i\|^{k-1}}\;;
    \quad1\le k\le d+1\ .
\label{eq:crit:deriv-bound}
\end{equation}
Note that this holds not
only inside $B_{\eps_i}(\novec\zeta_i)$ but in the whole domain of $f_i$, as
outside $B_{\eps_i}(\novec\zeta_i)$, $f_i$ is $\mathscr C^{d+1}$ and thus all
the derivatives are bounded.

Eq.~(\ref{eq:crit:deriv-bound}) is the key result for the remaining part of
the proof. First, it can be used to bound the integrand in
(\ref{eq:crit:int-after-partial}) by an integrable singularity (this is
most easily seen in spherical coordinates, where $1/r^{d-1}$ is integrable
in a $d$-dimensional space). Hence, the order of integration in
(\ref{eq:crit:int-after-partial}) can be interchanged, and it remains to
investigate the asymptotics of the integral
\begin{eqnarray}
I_i &=&\frac{1}{\|\ld\|^{d}}
    \int\limits_%
    {\makebox[1em][c]{\footnotesize$\zeta_{i,1}-r_i$}}^%
    {\makebox[1em][c]{\footnotesize$\zeta_{i,1}+r_i$}}
        \dd \phi_1
        g_i(\phi_1)
        \cos[\|\ld\|\phi_1-d\pi/2]\;,\quad \text{with}
\label{eq:crit:asympt-marginal}\\
g_i(\phi_1)&\equiv&\int\limits_%
    {\makebox[1em][c]{\footnotesize$B_{r_i}(\novec{}\tilde\zeta_i)$}}
    \dd^{d-1}\tilde\phi\;
    \tilde f_i^{(d)}(\phi_1,\novec{}\tilde\phi)\ .
\label{eq:crit:g-definition}
\end{eqnarray}
From (\ref{eq:crit:deriv-bound}), we now derive bounds on $g_i(\phi_1)$
and its first derivative. Again, we may safely fix $\phi_1\ne\zeta_{i,1}$ as this
has measure zero. Then, using (\ref{eq:crit:deriv-bound}) we find that
$$
|g_i(\phi_1)|\le\int_{0}^{r_i}\frac{C}{
    \left((\phi_1-\zeta_{i,1})^2+r^2)\right)^{(d-1)/2}}S_{d-1}r^{d-2}\dd r
$$
where we have transformed into spherical coordinates [$S_{d-1}$ is
the surface of the $(d-1)$-dimensional unit sphere] and assumed
the $l_2$-norm. Since $(\phi_1-\zeta_1)^2+r^2\ge r^2$, the
integrand can be bounded once again, and we find
\begin{eqnarray}
|g_i(\phi_1)|&\le&\int_{0}^{r_i}\frac{CS_{d-1}}{
    ((\phi_1-\zeta_{i,1})^2+r^2)^{1/2}}\dd r\nonumber\\
&=&
    C\left(-\log|\phi_1-\zeta_{i,1}|+
    \log\left[r_i+\sqrt{r_i^2+(\phi_1-\zeta_{i,1})^2}\right]\right)\nonumber\\
&\le&
-C\log|\phi_1-\zeta_{i,1}|
    \label{eq:crit:bound-on-g}
\end{eqnarray}
where in the last step we used that in
(\ref{eq:crit:asympt-marginal}) $|\phi_1-\zeta_{i,1}|<r_i$ and
that $r_i$ can be chosen sufficiently small.

Next, we derive a bound on $g_i'(\phi_1)$. As we fix $\phi_1\ne\zeta_1$,
the integrand in (\ref{eq:crit:g-definition}) is $\C^1$ and we can
take the differentiation into the integral,
$$
g'_i(\phi_1)=\int\limits_%
    {\makebox[1em][c]{\footnotesize$B_{r_i}(\novec{}\tilde\zeta_i)$}}
    \dd^{d-1} \novec{}\tilde\phi
    \tilde f_i^{(d+1)}(\phi_1,\novec{}\tilde\phi)\ .
$$
Again, we bound the integrand by virtue of
Eq.~(\ref{eq:crit:deriv-bound}) and obtain
\begin{eqnarray}
|g'_i(\phi_1)|&\le&\int_{0}^{r_i}\frac{CS_{d-1}}{
    ((\phi_1-\zeta_{i,1})^2+r^2)}\dd r\nonumber\\
&=&C\frac{\arctan\left[\frac{r_i}{|\phi_1-\zeta_{i,1}|}\right]}%
      {|\phi_1-\zeta_{i,1}|}\ \le\ \frac{C'}{|\phi_1-\zeta_{i,1}|}\ .
\label{eq:crit:bound-on-gprime}
\end{eqnarray}

Finally, these two bounds will allow us to estimate
(\ref{eq:crit:asympt-marginal}) and thus the asymptotics of the
correlations in the  lattice. We consider one half of the integral
(\ref{eq:crit:asympt-marginal}),
\begin{equation}
\int\limits_%
    {\makebox[1em][c]{\footnotesize$\zeta_{i,1}$}}^%
    {\makebox[1em][c]{\footnotesize$\zeta_{i,1}+r_i$}}
        \dd \phi_1
        g_i(\phi_1)
        \cos[\|\ld\|\phi_1-d\pi/2]\ ,
\label{eq:crit:g-fourier-int}
\end{equation}
as both halves contribute equally to the asymptotics. We then split the
integral at $\zeta_{i,1}+r_i/\|\ld\|$. The left part gives
\begin{eqnarray}
   \left|\int\limits_%
    {\zeta_{i,1}}^%
    {\zeta_{i,1}+r_i/\|\ld\|}
        \dd \phi_1
        g_i(\phi_1)
        \cos[\|\ld\|\phi_1-d\pi/2]\right|
&\stackrel{(\ref{eq:crit:bound-on-g})}{\le}&
    \int\limits_%
    {\zeta_{i,1}}^%
    {\zeta_{i,1}+r_i/\|\ld\|}
        \dd \phi_1(-\log|\phi_1-\zeta_{i,1}|)\nonumber\\
&=&\frac{r_i-r_i\log r_i+r_i\log \|\ld\|}{\|\ld\|}\ .
\label{eq:crit:left-int-asympt}
\end{eqnarray}
The right part of the split integral (\ref{eq:crit:g-fourier-int}) can be
estimated by integration by parts,
\begin{eqnarray}
&&\left|\int\limits_%
    {\zeta_{i,1}+r_i/\|\ld\|}^%
    {\zeta_{i,1}+r_i}
        \dd \phi_1
        g_i(\phi_1)
        \cos[\|\ld\|\phi_1-d\pi/2]\right|\le\nonumber\\
&&\qquad\le
    \left|\left[g_i(\phi_1)\frac{1}{\|\ld\|}\cos[\|\ld\|\phi_1-(d+1)\pi/2]\right]_%
    {\zeta_{i,1}+r_i/\|\ld\|}^{\zeta_{i,1}+r_i}\right|
    +
   \frac{1}{\|n\|}\hspace*{-1.5em}\int\limits_%
    {\zeta_{i,1}+r_i/\|\ld\|}^%
    {\zeta_{i,1}+r_i}\hspace*{-1.5em}
        \dd \phi_1
        |g'_i(\phi_1)|\nonumber\\[1ex]
&&\quad\ 
\stackrel{(\ref{eq:crit:bound-on-g},\ref{eq:crit:bound-on-gprime})}{\le}
C\frac{\log \|\ld\|}{\|\ld\|}+C'\frac{|\log r_i|}{\|\ld\|}\ .
\label{eq:crit:right-int-asympt}
\end{eqnarray}
Thus, both halves
[Eqs.~(\ref{eq:crit:left-int-asympt}),(\ref{eq:crit:right-int-asympt})]
give  a $\log \|\ld\|/\|\ld\|$ bound for the integral
(\ref{eq:crit:g-fourier-int}), and thus the integral
$I_i$ is asymptotically bounded by $\log \|\ld\|/\|\ld\|^{d+1}$ following
Eq.~(\ref{eq:crit:asympt-marginal}). As the number of such integrals in
(\ref{eq:crit:corr-element-sum}) is finite, this proves that
the correlations of the ground state decay at least as
$\log \|\ld\|/\|\ld\|^{d+1}$.
\hspace*{\fill}\qed
\end{proof}

\section{Gaussian Matrix Product States
    \label{sec:GaussianMPS}}

\begin{figure}[b]
\includegraphics[height=3cm]{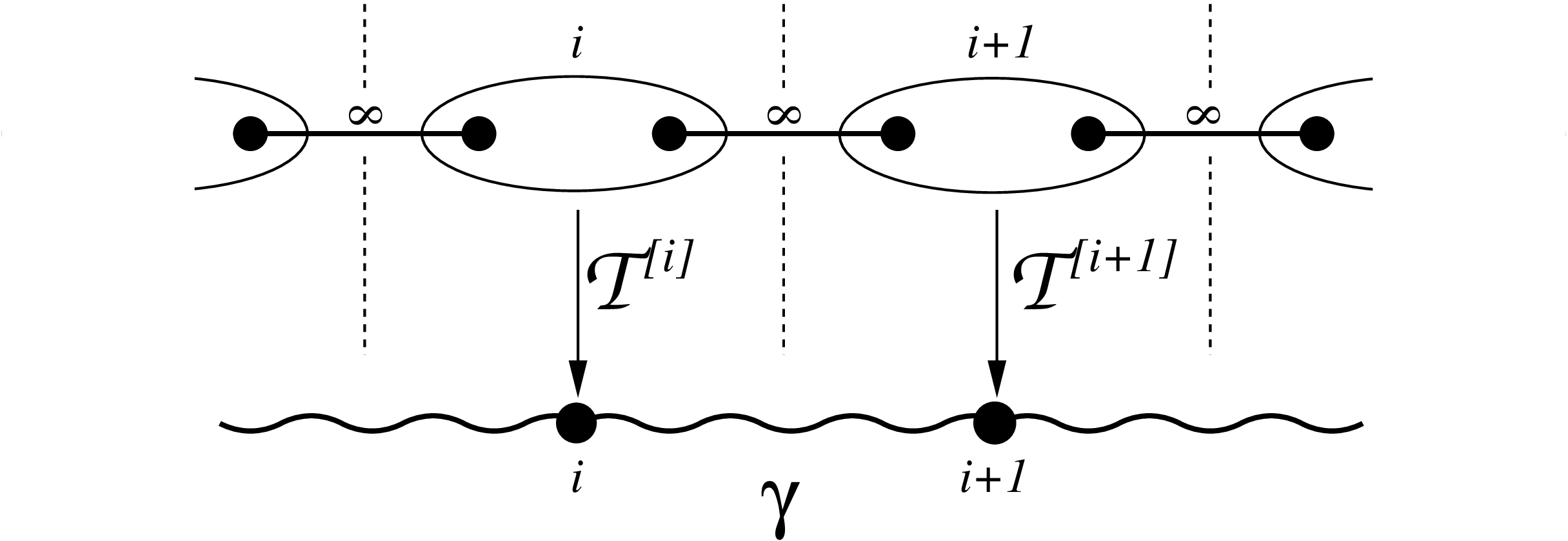}
\caption{
\label{fig:GMPS:operational-def}
Construction of Gaussian Matrix Product States (GMPS). GMPS are
obtained by taking a fixed number $M$ of maximally entangled (i.e., EPR) 
pairs shared by adjacent sites, and applying an arbitrary 
$2M$ to $1$ mode Gaussian operation $\mc T^{[i]}$ on site $i$.}
\end{figure}

Recently, so-called Matrix Product States (MPS) have attracted growing
interest in quantum information theory. These states appear in the DMRG
(Density Matrix Renormalization Group) method which is a powerful tool to
compute ground state properties of translational invariant Hamiltonians.
From a
quantum information perspective, this class of states can be given a
physical interpretation in terms of projected entangled pairs: they can be
obtained by taking a chain of maximally entangled pairs of dimension $D$
and applying a map $\mc T$ as in 
Fig.~\ref{fig:GMPS:operational-def} in a translational
invariant fashion. In the limit of large bond dimension $D$,
this allows to approximate arbitrary translational invariant states.
In finite dimensions, the MPS representation turned out to be a very
fruitful concept as it led to new powerful numerical algorithms~%
\cite{VGC04,ZV04,VC04,VPC04} accompanied by a better understanding of
their efficiency~\cite{VC05}, and new insights into renormalization group
transformations~\cite{VCL05} and sequential quantum
generators~\cite{SSV05}. In the following, we generalize matrix product
states to the Gaussian scenario.

\subsection{Definition of Gaussian MPS
    \label{sec:MPS:definition}}

We start by defining Gaussian matrix product states (GMPS). The definition
resembles the physical interpretation of finite-dimensional matrix product
states as projected entangled pairs (PEPs). In finite dimensions, MPS can
be described by taking maximally entangled pairs of dimension $D$ between
adjacent sites, and applying arbitrary local operations on each site,
mapping the $D\times D$ input to a $d$-dimensional output state.
Similarly, GMPS are obtained by taking a number of entangled bonds and
applying local (not necessarily trace-preserving) operations $\mc
T^{[i]}$, where  the boundary conditions can be taken either open or
closed.  Any GMPS is completely described by the type of
the bonds and by the operations $\mc T^{[i]}$.  Note that this
construction holds independent of the spatial dimension. For one
dimension, it is illustrated in Fig.~\ref{fig:GMPS:operational-def}.
As matrix product states are frequently used to describe translationally
invariant systems, an inportant case is given if all maps are identical,
$\mc T^{[i]}=\mc T\ \forall i$.

In order to define MPS in the Gaussian world, we have to
decide on the type of the bonds as well as on the type of operations.
We choose both the bonds to be Gaussian states and the operations to
be Gaussian operations, i.e., operations mapping Gaussian inputs to
Gaussian outputs.  For now, we will take the bonds to be maximally
entangled (i.e., EPR) states, such that the only parameter 
originating from the bonds is the number $M$ of EPRs.  
We show later on how the case of finitely entangled bonds can be easily
embedded.

\begin{figure}
\includegraphics[height=2cm]{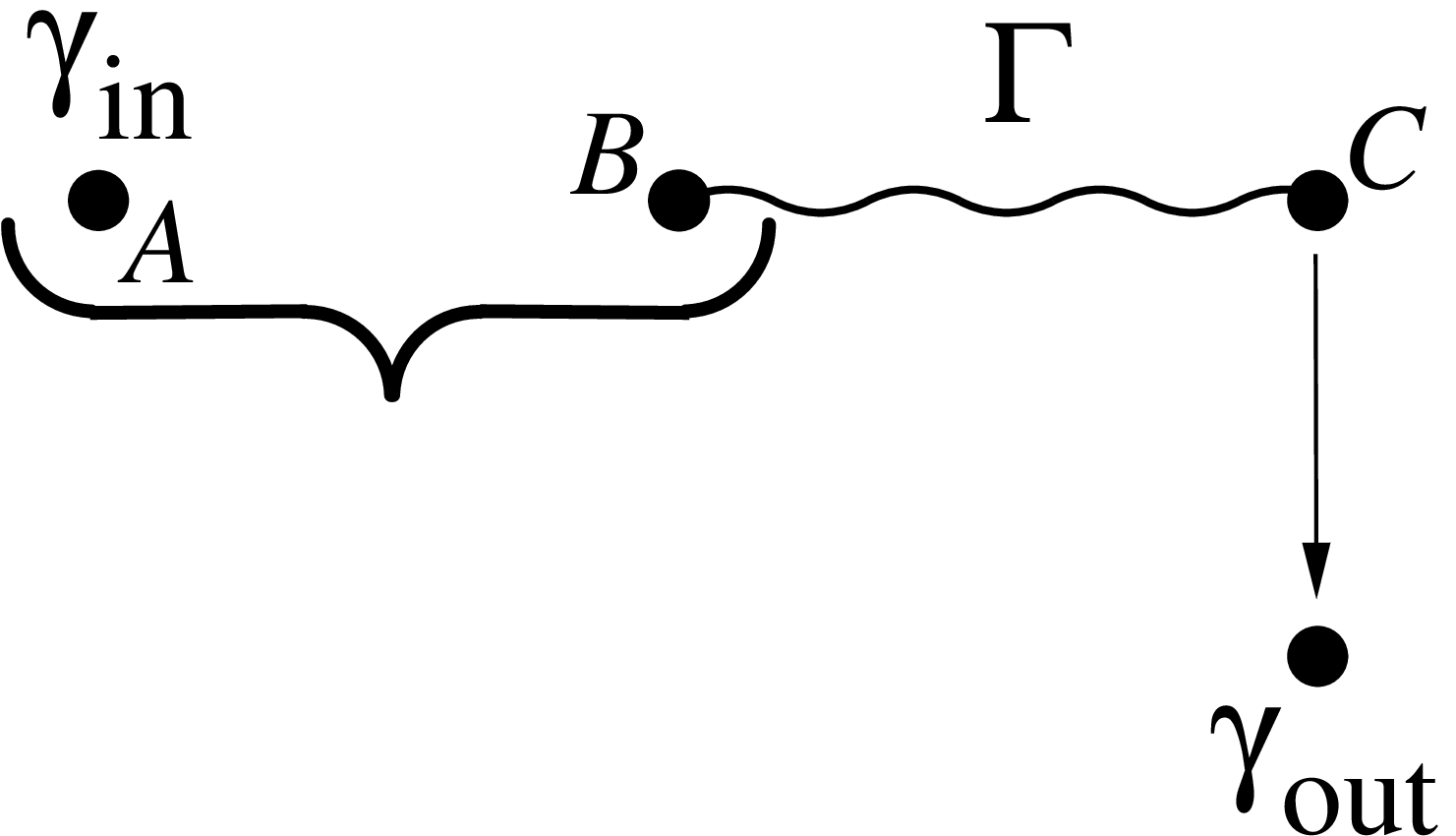}
\caption{
\label{fig:GMPS:jamiolkowski}
The Jamiolkowski isomorphism. The Gaussian channel described by
the state $\Gamma$ can be implemented by projecting the input state
$\gamma_\mathrm{in}$ (mode $A$) and the input port of $\Gamma$ (mode $B$) 
onto the EPR state (symbolized by curly brackets). In case of success, the
output is obtained in mode $C$.  The operation can be made trace-preserving
by measuring in a basis of displaced EPR states, and displacing $C$
according to the measurement outcome.
}
\end{figure}

As to the operations, we will allow for arbitrary Gaussian operations.
Operations of this type are most easily described by the Jamiolkowski
isomorphism~\cite{Jam72}.  There, any Gaussian operation $\mc T$ which maps
$N$ input modes to $M$ output modes can be described by an $N+M$ mode
covariance matrix $\Gamma$ with block $B$ (input) and $C$ (output).  The
corresponding map on some input state $\gamma_{\mathrm{in}}$ in mode $A$
is implemented by projecting the modes $A$ and $B$ onto an EPR state as
shown in Fig.~\ref{fig:GMPS:jamiolkowski}, such that the output state $\mc
T(\gamma_{\mathrm{in}})$ is obtained in mode $C$.  Conversely, the
matrix $\Gamma$ which represents the channel $\mc T$ is obtained by
applying the channel to one half of a maximally entangled state. The
duality between $\mc T$ and $\Gamma$  is most easily understood in terms
of teleportation, and shows that this characterization encompasses all
Gaussian operations. Note that the protocol of
Fig.~\ref{fig:GMPS:jamiolkowski} can be always made trace-preserving by
projecting onto the set of phase-space displaced EPR states and correcting the
displacement of mode $C$ according to the measurement outcome~\cite{GC02}.

In the following, we will denote all maps $\mc T$ by their corresponding
CM $\Gamma$.  Sometimes, we will speak of the modes $B$ and $C$ as input
and output ports of $\Gamma$, respectively.

We now discuss how the covariance matrix of the output will
depend on the CM of the input and on the channel
$\Gamma$~\cite{GC02,Fiu02b}.  This is most easily computed in the
framework of characteristic functions~\cite{Hol82}.
The characteristic function of the output is given by
$$
\chi_C(\xi_C)\propto\int e^{-\xi_A^T\gamma_{\mathrm{in}}\xi_A}
    e^{-\xi_{BC}^T\Gamma\xi_{BC}}
    \delta(x_A-x_B)\delta(p_A+p_B)\mathrm{d}\xi_{AB}\ ,
$$
and by integrating out subsystem $A$, 
$$
\chi_C(\xi_C)\propto\int e^{-\xi_{BC}^T M \xi_{BC}}\mathrm{d}\xi_B\ ,
$$
with
$$
M=\left(\begin{array}{cc}
    \theta\gamma\theta+\Gamma_B&\Gamma_{BC}\\\Gamma_{CB}&\Gamma_C
    \end{array}\right)\ .
$$
Basically, the integration
$\int\dd\xi_A\delta(x_A-x_B)\delta(p_A+p_B)$ does the following:
first, it applies the partial transposition 
$\theta\equiv\left(\begin{smallmatrix}1&0\\0&-1\end{smallmatrix}\right)$
to one of the subsystems, and second, it collapses the two systems $A$
and $B$ in the covariance matrix by adding the corresponding entries.  The
integration over $\xi_B$, one the other hand, leads to a state whose CM is
the Schur complement of $M_{11}$, $M_{22}-M_{21}M_{11}^{-1}M_{12}$, such
that the output state is described by the CM
$$
\gamma_{\mathrm{out}}=\Gamma_C-
    \Gamma_{CB}\frac{1}{\Gamma_B+\theta\gamma_{\mathrm{in}}\theta}\Gamma_{BC}\ .
$$

Let us briefly summarize how to perform projective measurements onto the
EPR state in the framework of CMs, where we denote the measured modes by $A$
and $B$, while $C$ is the remaining part of the system. First, apply the
partial transposition to $B$, second, collapse $A$ and $B$, and third,
take the Schur complement of the collapsed mode $AB$, which gives the
output CM of $C$.

As we discuss Gaussian matrix product states in connection with ground
states of Hamiltonians, we are mainly interested in pure GMPS.
Particularly, a GMPS is pure if the $\Gamma^{[i]}$ which describe the
operations $\mc T^{[i]}$ are taken to be pure, which we assume 
from now on.

Let us finally emphasize that the given defintion of MPS holds independent
of the spatial dimension of the system, as do most of the following
results, and in fact applies to an arbitrary graph.

\subsection{Completeness of Gaussian MPS\label{sec:MPS:completeness}}

In the following, we show that any pure and translational invariant
state can be approximated arbitrarily well by translational invariant
Gaussian matrix product states, i.e., GMPS with identical local operations
$\mc T$.  (Without translational invariance, this is clear anyway:
the complete state is prepared locally and teleported to its
destination using the bonds.) The proof is presented for one dimension,
but can be extended to higher spatial dimensions.
For the proof, we use a theorem on the simulation of translational
invariant Hamiltonians which is proven in the Appendix
(Theorem~\ref{theorem:app:simulation-of-interactions}). For our purposes, 
it says that in order to simulate an arbitrary translational invariant
Hamiltionian with reflection symmetry, it suffices if one can implement
translational invariant local and nearest neighbor Hamiltonians.%
\footnote{This naturally extends to higher dimensions. For two dimensions,
e.g., one needs nearest neighbor interactions and in addition one
interaction with the closest neighbor along the diagonal in order to break
the reflection symmetry. The implementation of such an interaction in the
MPS formalism is a straightforward extension of the presented method.}

\begin{figure}[b]
\includegraphics[height=7cm]{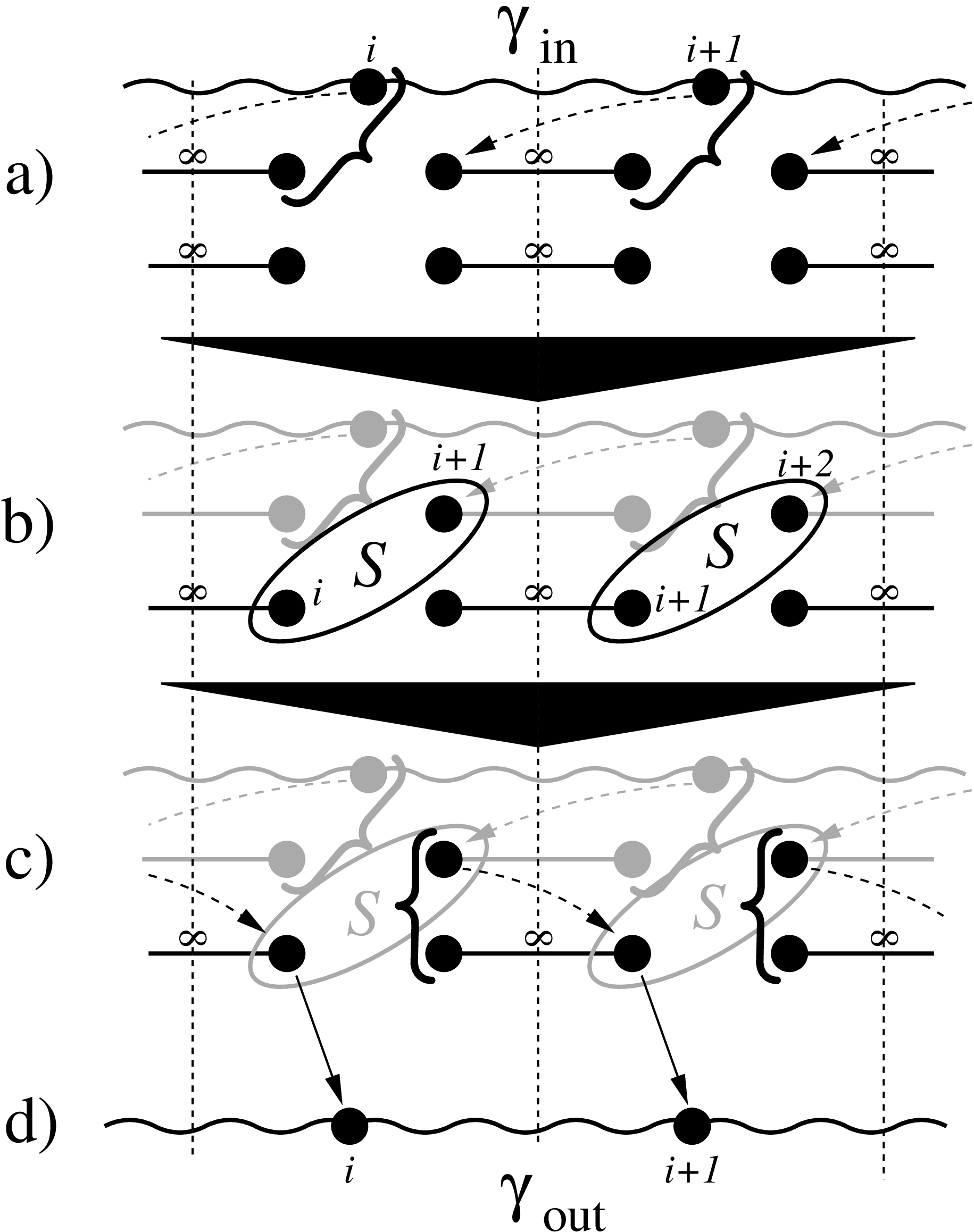}
\caption{
\label{fig:GMPS:completeness}
Implementation of a translational invariant nearest neighbor Hamiltonian
in a translational invariant fashion. Starting from $\gamma_\mathrm{in}$, 
the input ist first teleported to the left, then, the infinitesimal time
evolution $S=e^{\sigma H}$, $H\ll1$, is performed, and finally, the state
is teleported back.
}
\end{figure}

Given a translational invariant state $\gamma$, there is a
translational invariant Hamiltonian $H$ which transforms the separable
state $\openone$ into $\gamma$, $\gamma=SS^T$, $S=e^{\sigma H}$. According
to Theorem~\ref{theorem:app:simulation-of-interactions}, this time
evolution can be approximated arbitrarily well by a sequence of
translational invariant local (one-mode) and nearest neighbor (two-mode)
Hamiltonians $H_j$,
\begin{equation}
\label{eq:GMPS:trotter-decomp}
e^{\sigma H}\approx\prod_{j=1}^J e^{\bigoplus_n\sigma H_j}\ ,
\end{equation}
where the $H_j$ act on one or two modes, respectively, and approach the
identity for growing $J$.

Clearly, translational invariant local Hamiltionians can be implemented by
local maps without using any EPR bonds. In the following, we show how
translational invariant nearest-neighbor interactions can be implemented
by exploiting the entanglement of the bonds. The whole procedure is
illustrated in Fig.~\ref{fig:GMPS:completeness} and requires two EPR
pairs per site. We start with some initial state $\gamma_\mathrm{in}$ onto
which we want to apply $S_\oplus=e^{\bigoplus\sigma H_j}
    \approx\openone+\bigoplus_n\sigma H_j$.

First, we perform local EPR measurements between the modes of
$\gamma_\mathrm{in}$ and one of the bonds in order to teleport the modes
of $\gamma_\mathrm{in}$ to the left,
cf.~Fig.~\ref{fig:GMPS:completeness}a.  Then, the infinitesimal symplectic
operation $S=e^{\sigma H_j}$ is applied to the left-teleported mode and
the
second bond, Fig.~\ref{fig:GMPS:completeness}b.  In the last step,
another EPR measurement is performed which teleports the left-teleported
mode back to the right, and ``into'' the mode on which the adjacent $S$
was applied.  As the operations $e^{\sigma H_j}\approx\openone+\sigma H_j$
all commute, the ``nested'' application of the nearest neighbor symplectic
operations $S$ indeed give $S_\oplus$.  Thus, the remaining mode indeed
contains the output
$\gamma_\mathrm{out}=S_\oplus\gamma_\mathrm{in}S_\oplus^T$.  The whole
decomposition (\ref{eq:GMPS:trotter-decomp}) can be implemented by
iterated application of the whole protocol of Fig.~\ref{fig:GMPS:completeness}.

\subsection{Gaussian MPS with finitely entangled bonds}

\begin{figure}[b]
\includegraphics[width=.95\textwidth]{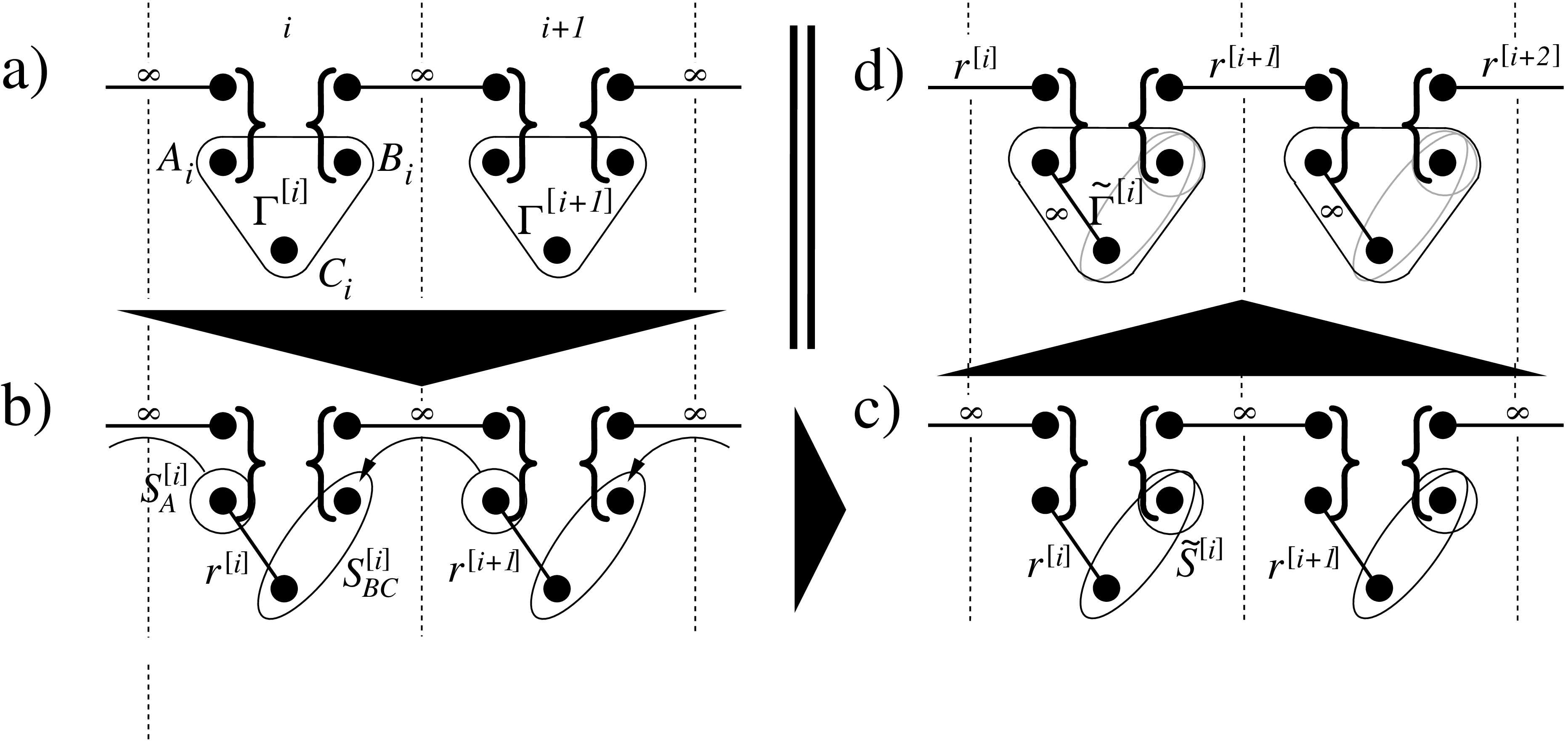}
\caption{
\label{fig:GMPS:nonmax-ent}
How to make the bonds of GMPS finitely entangled. \textbf{a)} The initial
MPS.  \textbf{b)} Do a Schmidt decomposition of the original map $\Gamma$.
\textbf{c)} Move the $S^{[i]}_A$ through the infinitely entangled bond to
the next site.  \textbf{d)} Swap the finitely and the infinitely entangled
pair.
}
\end{figure}

In this subsection we show that in general, infinitely entangled
bonds can be replaced by finitely entangled ones.  Intuitively, this
should be possible whenever the channel $\mc T^{[i]}$ destroys some
of the entanglement of the bond anyway, i.e., $\Gamma^{[i]}$ is non-maximally
entangled. In that case, it should be possible to use a less entangled
bond while choosing a channel which does not destroy entanglement any
more.

The method is illustrated in Fig.~\ref{fig:GMPS:nonmax-ent}. Again, for
reasons of clarity we restrict to one dimension and one bond. The
argument however appies independent of the spatial dimension and the
number of bonds. The only restriction we have to make is the restriction
to pure GMPS, i.e., those with pure $\Gamma^{[i]}$.

Consider a GMPS with local channels given by $\Gamma^{[i]}$ and infinitely
entangled bonds, Fig.~\ref{fig:GMPS:nonmax-ent}a. First, apply a Schmidt
decomposition~\cite{HW01} to $\Gamma^{[i]}$ in the partition $A|BC$, which
can be always done as long as $\Gamma^{[i]}$ is pure. The Schmidt
decomposition allows us to rewrite the state as shown in
Fig.~\ref{fig:GMPS:nonmax-ent}b---an entangled state between modes $A$ and
$C$ with two-mode squeezing $r^{[i]}$, $B$ in the coherent state
$\openone$, and sympectic operations $S_A^{[i]}$ and $S_{BC}^{[i]}$ which
are applied to modes $A$ and $BC$, respectively. As the bond itself is
infinitely entangled, we can teleport the sympectic operation through the
bond to the next site as
indicated in Fig.~\ref{fig:GMPS:nonmax-ent}b. Then, $S_A^{[i+1]}$ can be
merged with $S_{BC}^{[i]}$ to a new operation $\tilde S^{[i]}$ acting on
modes $B$ and $C$ of site $i$ (Fig.~\ref{fig:GMPS:nonmax-ent}c). Finally, 
in the triples consisting of one maximally entangled state, one non-maximally
entangled state, and the projection onto the EPR state, the maximally and
the non-maximally entangled state can be swapped, resulting in
Fig.~\ref{fig:GMPS:nonmax-ent}d. There, we have finitely entangled bonds,
while the infinite entanglement has been moved into the new maps
$\tilde\Gamma^{[i]}$.

It is tempting to apply this construction to the completeness proof of the
preceding section in order to obtain a construction which is less wasting
with respect to resources. However, for any iterative protocol this is
most likely difficult to achieve. The reason for this is found in the
no-distillation theorem which states that with Gaussian operations, it is
not possible to increase the amount of entanglement~\cite{GC02} between two
parties.  Particularly, this implies that in each step of an iterative
protocol, the bonds need to have at least as much entanglement as can be
obtained at the output of this step, maximized over all inputs where the
entanglement is increased. This is indeed a severe restriction, although
it does not imply the impossibility of such a protocol. One could, e.g.,
create a hightly
entangled state in the first step and then approach the desired state by
decreasing the entanglement in each step. Still, it seems most likely that a
sequence of MPS which approach a given state efficiently will have to
involve more and more bonds simultaneously and thus cannot be constructed
in an iterative manner.

\subsection{Correlation functions of Gaussian MPS}

\begin{figure}[b]
\includegraphics[width=7cm]{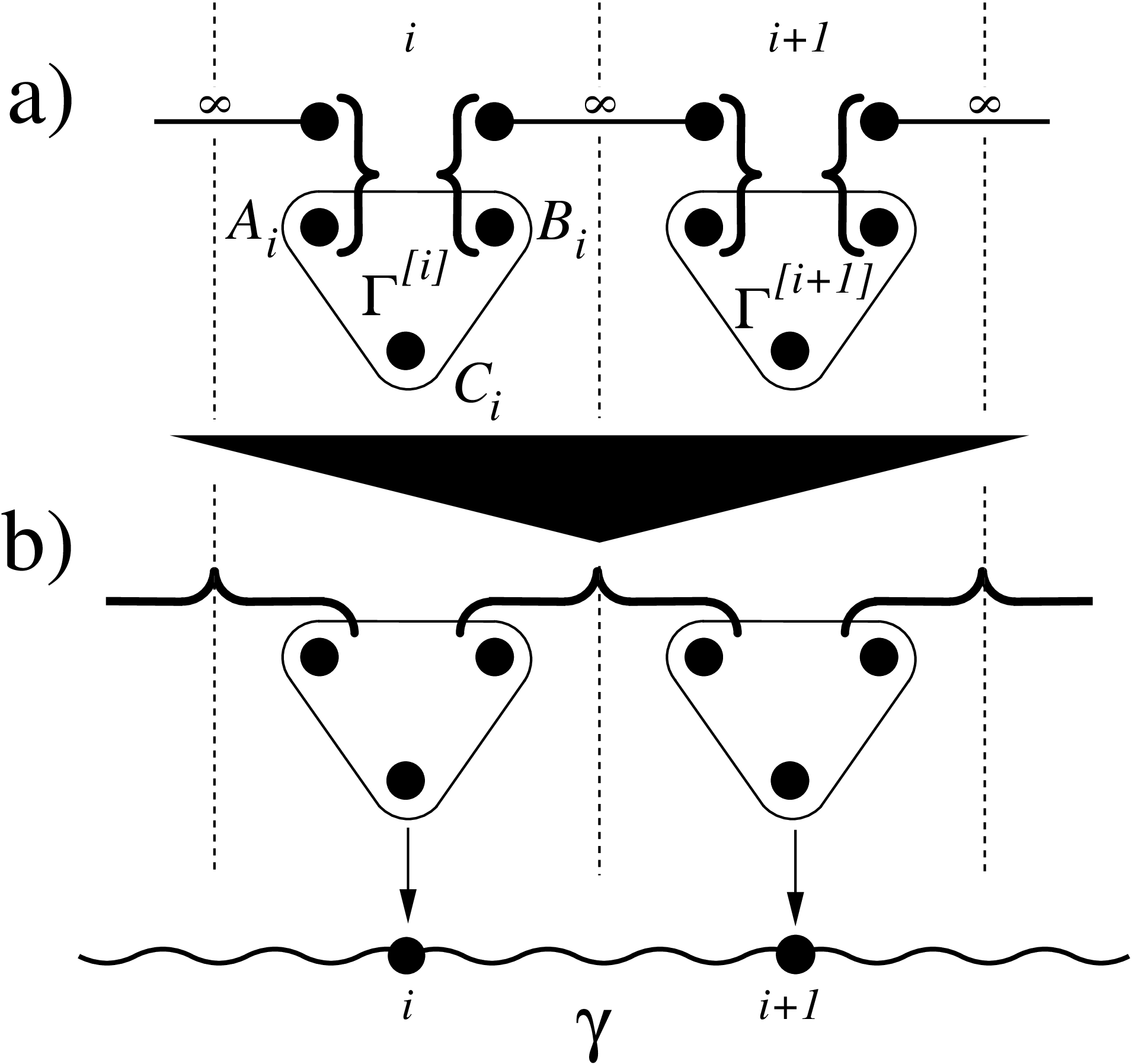}
\caption{If the local operations are described by states $\Gamma^{[i]}$ via
the Jamiolkowski isomorphism, the construction of GMPS can be simplified
by replacing the measurement-bond-measurement triples by a simple
projection onto the EPR state.
\label{fig:GMPS:mps-definition-with-jamiolkowski}
}
\end{figure}

In this section, we show how to compute correlations functions from the
maps $\Gamma^{[i]}$ which describe the GMPS. We show that this can be
done efficiently, i.e., in polynomial time independent of the dimension
of the graph which is different to the finite dimensional case. Of
course, this is not too surprising as Gaussian states can be fully
characterized by a number of paramaters quadratic in the number of modes.

Let us start with the general case of different $\Gamma^{[i]}$, as in
Fig.~\ref{fig:GMPS:mps-definition-with-jamiolkowski}a. The calculation can
be facilitated by the simple observation that the triples consisting of
two projective measurements and one EPR pair can be replaced by a single
projection onto the EPR state,
Fig.~\ref{fig:GMPS:mps-definition-with-jamiolkowski}b. It follows that we 
can apply the formalism for projective measurements onto the EPR
state which we presented in Sec.~\ref{sec:MPS:definition}. 
We start from $\bigoplus_i \Gamma^{[i]}$.
First we partially transpose all $A$ modes, then we collapse $A_{i+1}$ and
$B_i$ for all $i$, and finally we take the Schur complement of the merged
mode. In case of periodic boundary conditions, this can be expressed by
the transformation matrix 
\begin{equation}
\label{eq:GMPS:M-Matrix}
\Pi=\left(\begin{array}{ccc}
    \openone_A & \mathcal R\theta_B & 0 \\
    0 & 0 & \openone_C 
  \end{array}\right)
\end{equation}
which maps $ABC$ onto $A'C$, where $\theta_B\equiv\theta\otimes\openone$
is the partial transposition on system $B$, and $\mathcal R$ is the
circulant right shift operator, 
$(\mathcal R)_{ij}=\delta_{i,j+1\mod N}\otimes\openone$.  Then, the
output state, i.e., the GMPS characterized by $\Gamma^{[i]}$, is
$$
\gamma=\mathrm{SC}_{A'}\left[\Pi\left(
	    \bigoplus_i\Gamma^{[i]}\right)\Pi^T\right]\ ,
$$
where $\mathrm{SC}_{X}(U)$ is the Schur complement of the $X$ part of $U$, 
$\mathrm{SC}_{X}(U)=U_{YY}-U_{YX}U_{XX}^{-1}U_{XY}$. For fixed boundary
conditions, the matrix $\Pi$ has to be modified accordingly at the
boundaries. All the involved operations scale polynomially in the product
$NM$ of the number of sites $N$ and the number of modes $M$.

In case all the local maps are chosen equal,
$\Gamma^{[i]}\equiv\Gamma\ \forall i$, the above formula can be simplified
considerably. 
 Therefore, note that the Fourier transform can be taken into the
Schur complement, and that $\Pi$ as well as $\bigoplus_{i=1}^N
\Gamma^{[i]}=\Gamma\otimes\openone_N$ are blockwise circulant so that both are
diagonalized by the Fourier transform.  We again adapt the 
notation of writing the diagonal of the Fourier transformed matrices as
functions of an angle $\phi$, cf.~Sec.~\ref{sec:tinv-systems}. In that
case, $\Gamma\otimes\openone$ is mapped onto the constant function
$\Gamma$, and the same holds for
$\openone$ and $\theta$ in (\ref{eq:GMPS:M-Matrix}). The right shift
operator $\mc R$, on the other hand, is transformed to $e^{i\phi}$\openone: 
the EPR measurement performed between adjacent sites leads to a complex
phase of $\phi$. Altogether, we have
$$
\hat{\Pi}=\left(\begin{array}{ccc}
    \openone_A & e^{i\phi}\theta_B & 0 \\
    0 & 0 & \openone_C 
  \end{array}\right)\ ;\quad
\hat\gamma=\mathrm{SC}_{A'}\left[\hat{\Pi}\,\Gamma\,\hat{\Pi}^\dagger\right]\ .
$$
Directly expressed in terms of the map $\Gamma$, this reads
\begin{equation}
\hat\gamma(\phi)=\Gamma_C-\Gamma_{C|AB}\,\hat\Lambda
    \frac{1}{\hat\Lambda\,\Gamma_{AB|AB}\,\hat\Lambda^\dagger}\,
	\hat\Lambda^\dagger\,\Gamma_{AB|C}
\label{eq:MPS:characterization}
\end{equation}
where $\hat\Lambda=(\openone_A\,;\,e^{i\phi}\theta_B)$ is 
the upper left subblock of $\hat{\Pi}$.

\subsection{States with rational trigonometric functions as Fourier transforms}

If one restricts to pure MPS (i.e.,  those for which $\Gamma$ is
pure) with one mode per site,  then it
follows from Theorem~\ref{theorem:pure-reflection-symmetry} that these
states have reflection symmetry, and therefore
$\hat\gamma(\phi)=\gamma_0+2\sum_{n\ge0}\gamma_n\cos(\ld\phi)$ is real.
This implies that the sines in (\ref{eq:MPS:characterization}) can only
appear in even powers $\sin^{2n}\phi=(1-\cos^2\phi)^n$. Therefore, the
Fourier transform $\hat\gamma$ of any pure Gaussian MPS, which is a
$2\times 2$ matrix valued function of $\phi$, has elements which are
rational functons of $\cos(\phi)$,
$(\hat\gamma(\phi))_{xy}=p_{xy}(\cos(\phi))/q_{xy}(\cos(\phi))$ with $p$,
$q$ polynomials. The degree of the polynomials is limited by the 
size of $\hat\Lambda\Gamma_{AB}\hat\Lambda^\dagger$, and thus by the
number $M$ of the bonds. One can easily check that $\dim p\le2M+1$ and
$\dim q\le2M$.

For the following discussion, let us write those rational functions with a
common denominator $d$,
\begin{equation}
\label{sachertorte}
\hat\gamma(\phi)=\frac{1}{d(\cos(\phi))}
    \left(\begin{array}{cc}
	q(\cos(\phi)) & r(\cos(\phi)) \\ r(\cos(\phi)) & p(\cos(\phi)) 
    \end{array}\right)\ ,
\end{equation}
where $q$, $p$, $r$, and $d$ are polynomials of degree $L$. Then, 
the set of all such $\hat\gamma$ with $L\ge2M+1$ encompasses the set of
translational invariant GMPS with $M$ bonds. Computing correlation
functions in a lattice of size $N$ can be done straightforwardly in this
representation by taking the discrete Fourier transform of
$\hat\gamma(\phi)$ which scales polynomially with $N$, and in the
following section we show that for one dimension, the correlations can be
even computed exactly in the limit of an infinite chain.

It is interesting to note that $\gamma(\phi)$ is already determined 
up to a finite number of possibilities by fixing $r$ and $d$. Since
$\gamma$ is pure, $1=\det\gamma=\det\hat\gamma$, and therefore,
$pq=d^2+r^2$. Therefore, the zeros of $pq$ are the zeros of $d^2+r^2$, 
such that the only freedom is to choose how to distribute the zeros on $p$
and $q$. On the contrary, fixing only $q$ and $d$ does not give sufficient
information, while choosing $p$, $q$ and $d$ (i.e., the diagonal of
$\hat\gamma$) does not ensure that there
exists a polynomial $r$ such that $pq-r^2=d^2$.

From the above, it follows that $2L+1$ parameters are sufficient to
describe $\hat\gamma(\phi)$, where $L$ is still the degree of the
polynomials. This encloses all translational invariant Gaussian MPS with
bond number $M\le(L-1)/2$, which need $(2M+1)(2M+2)=L(L+1)$ parameters.
Therefore, the class of states where $\hat\gamma(\phi)$ is a rational
function of $\cos(\phi)$ is a more efficient description of
translationally invariant states than Gaussian MPS are.

Let us stress once more that the results of this section hold 
for arbitrary spatial dimension.

\subsection{Correlation length}

In the following, we show that the correlations of one-dimensional GMPS
decay exponentially and explicitly derive the correlation length.  The
derivation only makes use of the representation (\ref{sachertorte}) of
Gaussian MPS and thus holds for the whole class of states where the
Fourier transform is a rational function of the cosine. We will restrict
to the case where the state $\Gamma$ associated to the GMPS map  has only 
finite entries, which corresponds to the case where the denominator
$d(\cos(\phi))$ in (\ref{sachertorte}) has no zero on the unit circle.%
\footnote{
The case where $d$ has zeros on the unit circle corresponds to critical
systems, which is why the correlations diverge. In the case of a
Hamiltonian $H=V\oplus\openone$, however, the ground state correlations of
$P$ do not diverge. As in that case one has $pq=d^2$, 
$p/d=d/q$ need not have a singularity just because $q/d$ has one.}

The
correlations are directly obtained by back-transforming the elements of
$\hat\gamma(\phi)$, which are rational functions
$[\hat\gamma(\phi)]_s=s(\cos(\phi))/d(\cos(\phi))$, $s\in\{p,q,r\}$; 
in the limit of an infinite chain,
$$
(\gamma_s)_n=\frac{1}{2\pi}\int_0^{2\pi}
    \frac{s(\cos(\phi))}{d(\cos(\phi))}e^{i\ld\phi}\dd\phi\ .
$$
Now transform $s$, $d$ to complex polynomials via
$\cos(\phi)\rightarrow(z+1/z)/2$, and  expand with $z^K$, $\tilde
s(z):=z^Ks(z)$, $\tilde d(z):=z^Kd(z)$, where $K$ is chosen large
enough to make $\tilde s$, $\tilde d$ polynomials in $z$. Then,
\begin{eqnarray*}
(\gamma_s)_n&=&\frac{1}{2\pi i}\int_{\mc S^1}
    \frac{\tilde s(z)z^{\ld-1}}{\tilde d(z)}\dd z\\
&=&\sum_{z_i:\tilde d(z_i)=0}\frac{1}{(\nu_i-1)!}
    \underbrace{
    \left.\frac{\dd^{\nu_i-1}}{\dd z^{\nu_i-1}}
    \left[\frac{\tilde s(z)z^{\ld-1}}{\tilde d_i(z)}\right]
    \right|_{z=z_i}}_{D_i}
\end{eqnarray*}
by the calculus of residues, where $\nu_i$ is the order of the zero $z_i$
in $\tilde d$ and $\tilde d_i(z)(z-z_i)^{\nu_i}=\tilde d(z)$.  For
$\ld>\nu_i$, $D_i\propto z_i^{(\ld-\nu_i)}$, and it follows that the
correlations decay exponentially, where the correlation length is given
by the largest zero of $q(z)$ inside the unit circle.

This proof holds only for one-dimensional GMPS. However, it can be proven
for arbitrary spatial dimensions that the correlations decay as
$o(\|n\|^{-\infty})$ by iterated integration by parts as in
Lemma~\ref{lemma:rapid-H-rapid-E}.

\subsection{GMPS as ground states of local Hamiltonians}

Finally, we prove that every Gaussian MPS is the ground state of a
local Hamiltonian, and show that only a special class of local
Hamiltonians has a GMPS as an exact ground state. Again, the proof only relies
on the representation (\ref{sachertorte}). 

Given a state $\gamma$ with Fourier transform (\ref{sachertorte}), define
$H$ to be the Hamiltonian matrix with Fourier transform
\begin{equation}
\label{eq:GMPS:local-hamil-gs}
\hat H(\phi)=
    \left(\begin{array}{cc}
	p(\cos(\phi)) & -r(\cos(\phi)) \\ -r(\cos(\phi)) & q(\cos(\phi)) 
    \end{array}\right)\ .
\end{equation}
Then, $H$ corresponds to a local Hamiltonian---the interaction range is
the degree of $p,q,r$---%
and according to (\ref{eq:basics:tinv-Ehat-definition}), 
$\E(\phi)=\big[\sqrt{pq-r^2}\;\big](\cos\phi)=d(\cos\phi)$, which together
with Eq.~(\ref{eq:basics:tinv-E0-and-gamma_from_Ehat}) proves that 
$\gamma$ is the ground state of $H$.

It is interesting to have a brief look at the converse as well. Given a local
Hamiltonian, when will it have a GMPS as its ground state?
Any local Hamiltonian has a Fourier transform which consists of
polynomials in $\cos(\phi)$, and thus we adapt the notation of
Eq.~(\ref{eq:GMPS:local-hamil-gs}). Then, the ground state is represented
by a rational function of $\cos(\phi)$ in Fourier space exactly if 
$pq-r^2=d^2$ is the square of another polynomial, as can be seen from
Eq.~(\ref{eq:basics:tinv-E0-and-gamma_from_Ehat}). In terms of the
original Hamiltonian, this implies that
$H_QH_P-H_{QP}^2$ has to be the square of another banded matrix. 
For example, for the usual case $H=V\oplus\openone$ one would need
$V=X^2$ with $X$ again a banded matrix.  The Klein-Gordon
Hamiltonian (\ref{eq:basics:kleingordon}), e.g., does not have a GMPS as
its ground state.

\acknowledgements

We would like to thank Jens Eisert, Otfried G\"uhne, David P\'erez
Garc\'\i a, Diego \mbox{Porras}, Tommaso Roscilde, Frank Verstraete, and Karl
Gerd Vollbrecht for helpful discussions and comments. This work has been
supported by the EU IST projects QUPRODIS and \mbox{COVAQIAL}.

\appendix

\section{Hamiltonian simulation\
    \label{sec:appendix:ham-simul}}

In this Appendix, we discuss the following question.
Consider the set $\cal S$ of symplectic
transformations of $N$ harmonic oscillators on a ring which have
both translation and reflection symmetry. Assume that we can
implement every local transformation of the form
$S\oplus\ldots\oplus S \in {\cal S}$ and in addition one element
of $\cal S$ corresponding to a nearest-neighbor interaction
Hamiltonian.  Is this set universal for simulating any operation in
$\cal S$?

\subsection{Preliminaries}

\textbf{Lie-Trotter formulae:} It is most convenient to
discuss the problem on the level of the Lie algebra of the
symplectic group, since starting with a fixed set of operations in
$\cal S$ the ``reachable'' transformations are characterized by
the closure of the Lie algebra. That is, if  $e^{\lambda A}$ and
$e^{\lambda B}$ are reachable for every real $\lambda$ then so is
$e^{\alpha A+\beta B}$ and $e^{\gamma[A,B]}$ for all
$\alpha,\beta,\gamma\in \mathbb{R}$. This follows from the
{\it{Lie-Trotter formulae}}
\begin{eqnarray}
e^{\alpha A+\beta B}&=&
    \lim_{n\rightarrow\infty}\left(e^{\alpha A/n}
    e^{\beta B/n}\right)^n \mbox{ and }
    \label{Eq:Trotter1}
\\
e^{[A,B]}&=&
\lim_{n\rightarrow\infty}\left(e^{ A/\sqrt{n}}e^{ B/\sqrt{n}}e^{-
A/\sqrt{n}}e^{- B/\sqrt{n}}\right)^n\ .
\label{Eq:Trotter2}
\end{eqnarray}

\textbf{The Lie algebra:} 
Let $S=e^{t A}$ be a symplectic transformation close to the
identity, i.e., $|t|\ll 1$. By examining the first order in $t$ we
see that the condition $S\sigma S^T=\sigma$ is equivalent to
$A\sigma=(A\sigma)^T$. Hence, we get all possible generators $A$
by multiplying any real symmetric matrix with the symplectic
matrix $\sigma$ so that \begin{equation}\label{Eq:Agen}
A=\left(\begin{array}{cc}
  P & C \\
  C^T & Q \\
\end{array}\right)\sigma^T=\left(\begin{array}{cc}
  -C & P \\
  -Q & C^T \\
\end{array}\right),
\end{equation} where $Q,P$ are symmetric and the block structure
corresponds to a direct sum of momentum and position space.

If $S\in {\cal S}$ then $Q,P$ and $C$ have to be circulant
matrices (translation symmetry) and $C=C^T$ (reflection symmetry).
Under these symmetries (and utilizing the fact that all circulant
matrices mutually commute) the blocks of the commutator
$A''=[A,A']$ are given by
\begin{eqnarray}
P''&=&2(C'P-CP'),\nonumber\\ Q''&=&2(CQ'-C'Q),\label{Eq:comblocks}\\
C''&=&Q'P-QP'\nonumber.
\end{eqnarray}
Note that, even if we do not assume reflection symmetry for $A$
and $A'$, then $A''$ will have it, i.e., $C''$ will be symmetric.
For this reason it is crucial to impose reflection symmetry --
otherwise the set of operations will not be universal.

For {\it local transformations} of the form $S\oplus\ldots\oplus S
\in {\cal S}$ the blocks $P,Q$ and $C$ are proportional to the
identity, since $\bigoplus_i e^{A_i}=e^{\bigoplus_i A_i}$.

\textbf{Quadratic Hamiltonians:} 
The relation between the Hamiltonian matrix of a quadratic
Hamiltonian and the generator of
the respective symplectic transformation can be obtained from the
equation
\begin{equation}
e^{i\mc {H}t} R_\ld e^{-i\mc {H}t} \equiv \sum_l S_{kl} R_l,\quad
S=e^{At}.
\end{equation}Examining the infinitesimal regime yields \begin{equation}
A=\sigma H=\left(\begin{array}{cc}
  H_{QP}^T & H_P \\
  -H_Q & -H_{QP} \\
\end{array}\right).
\end{equation}
Note that in general a symplectic transformation generated by an
$A$ of the form in Eq.~(\ref{Eq:Agen}) does not correspond to a
semi-bounded Hamiltonian. However, by Eq.~(\ref{Eq:Trotter1}) we
can always decompose any symplectic transformation into time
evolutions each of which is governed by a semi-definite
Hamiltonian matrix. Moreover, if we can simulate the time
evolution governed by a Hamiltonian $\mc  H$ we can in principle
also simulate the evolution according to $-\mc  H$ by going to the
revival time $T_\mathrm{rev}$ for which
\begin{equation}\label{Eq:Trev}
e^{i \mc {H} (T_\mathrm{rev}-t)}=e^{-i \mc {H} t}.
\end{equation}

\subsection{Universality}
\begin{theorem}
\label{theorem:app:simulation-of-interactions}
Consider any transformation 
corresponding to a nearest-neighbor interaction Hamiltonian
in the set $\cal S$ of translationally
invariant symplectic operations with reflection symmetry.
Together with all local transformations of the form $S\oplus\ldots\oplus S
\in {\cal S}$
this is universal for simulating any transformation in $\cal S$.
\end{theorem}

\begin{proof}
The theorem is proven in two steps. First we
show that using additional local transformations only, we can {\it
extract} each block in Eq.~(\ref{Eq:Agen}) in the sense that one
can simulate a new operation, where two of the three matrices
$Q,P,C$ are set to zero, while the third remains unchanged. In
this way we can in the second step discuss the reachability for
each of these three
components separately:

\textit{Extracting the blocks.}---We will only show how the
$P$-block can be extracted. Similar reasoning will then apply for
the $Q$-block and by utilizing Eq.~(\ref{Eq:Trotter1}) we can
extract the $C$-components by simply subtracting the $Q$ and $P$
blocks.

Let us start with a general $A$ of the form in Eq.~(\ref{Eq:Agen})
and first get rid of the $C$-component. To this end consider the
commutation relation (\ref{Eq:comblocks}) with $C'=\frac12\openone$
and $Q'=P'=0$. Then $P''=P, Q''=Q$ and $C''=0$. Starting with an
$A$ of this form ($C=0$) we may take the commutator with
$Q'=\frac12{\bf 1}, C'=P'=0$. Then $P''=Q''=0$ and $C''=P$. The
commutator of such a matrix in turn with $P'=-\frac12\openone$,
$Q'=C'=0$ leads then finally to $P''=P$ with the other components
zero.

\textit{Generating a basis.}---%
Consider a matrix $A$ which corresponds to a nearest-neighbor
interaction, i.e. in one of the blocks $Q,P,C$ the first
off-diagonal is non-zero. By taking commutators like in (i) we can
easily construct a second generator $A'$, also corresponding to
such an interaction, however with the respective off-diagonal in
one of the other two blocks. By Eq.~(\ref{Eq:comblocks}) taking the
commutator $[A,A']$ will lead to a product of the two non-diagonal
circulant matrices in one of the blocks. This product will again
be a circulant matrix, but now with a non-zero second
off-diagonal. By iterating this procedure we can subsequently
generate non-zero entries in every off-diagonal and taking linear
combinations of these matrices will thus lead to a basis of
symmetric circulant matrices in each of the blocks. Hence,
together with (i) every generator corresponding to an element of
$\cal S$ can be constructed.
\hspace*{\fill}\qed
\end{proof}

\subsection{Remarks}
\textbf{Revival time and efficiency:} Making use of the
    revival time (\ref{Eq:Trev}) might be a severe handicap
    concerning the efficiency of the simulation. In particular
    $T_\mathrm{rev}$ of the interaction Hamiltonian $H_\mathrm{int}$ might grow
    exponentially with the number $N$ of modes.

    In this case there are two ways to speed up the simulation:
    either one supplements the set of transformations with an
    additional interaction which is such that it provides an efficient
    simulation of $-H_\mathrm{int}$, or one starts with a ``good''
    $H_\mathrm{int}$ for
    which this is possible from the very beginning. Examples for the
    latter are the Hamiltonians
    \begin{eqnarray}
    \mc {H}&=&\sum_{i=1}^N{Q}_i^2+{P}_i^2 + \alpha
    ({Q}_i{P}_{i+1}+{P}_i{Q}_{i+1}),\\
    \mc {H}&=&\sum_{i=1}^N({Q}_i+{Q}_{i+1})^2+
    ({P}_i-{P}_{i+1})^2.
    \end{eqnarray}
    In both cases we can efficiently simulate the evolution according to
    $-\mc {H}$ by first applying the symplectic transformation
    ${Q}\mapsto{P},{P}\mapsto-{Q}$ and then
    changing the sign of the diagonal in $H$ by local operations.

\end{document}